\documentclass[12pt,a4paper]{article}

\usepackage{epsfig}
\usepackage{psfig}
\usepackage{exscale}
\def\lsim{\raise0.3ex\hbox{$<$\kern-0.75em\raise-1.1ex\hbox{$\sim$}}}
\def\gsim{\raise0.3ex\hbox{$>$\kern-0.75em\raise-1.1ex\hbox{$\sim$}}}
\newcommand{\be}{\begin{equation}}
\newcommand{\ee}{\end{equation}}
\newcommand{\ba}{\begin{eqnarray}}
\newcommand{\ea}{\end{eqnarray}}

\def\spose#1{\hbox to 0pt{#1\hss}}
\def\ltapprox{\mathrel{\spose{\lower 3pt\hbox{$\mathchar"218$}}
 \raise 2.0pt\hbox{$\mathchar"13C$}}}
\def\gtapprox{\mathrel{\spose{\lower 3pt\hbox{$\mathchar"218$}}
 \raise 2.0pt\hbox{$\mathchar"13E$}}}
\def\phv{\vec \phi}

\def\ad#1{$\,^{\rm #1}$}
\def\NT{N_\tau}
\def\nt{\ifmmode\NT\else$\NT$\fi}
\def\NS{N_\sigma}
\def\ns{\ifmmode\NS\else$\NS$\fi}

\def\p{^\prime}

\def\n{\noindent}

\begin{document}
\begin{titlepage} 
\thispagestyle{empty}

 \mbox{} \hfill BI-TP 2001/21\\
 \mbox{} \hfill May 2002\\
 \mbox{} \hfill cond-mat/0202017
\begin{center}
\vspace*{1.0cm}
{{\Large \bf Universal amplitude ratios from numerical\\         
studies of the three-dimensional $O(2)$ model\\}}\vspace*{1.0cm}
{\large A. Cucchieri\ad a, J. Engels\ad b, S. Holtmann\ad b, T. Mendes\ad a,
 T. Schulze\ad b}\\ \vspace*{0.8cm}
\centerline {{\large $^{\rm a}$}{\em IFSC-USP, Caixa postal 369,
    13560-970 S\~ao Carlos SP, Brazil}} \vspace*{0.4cm}
\centerline {{\large $^{\rm b}$}{\em Fakult\"at f\"ur Physik, 
    Universit\"at Bielefeld, D-33615 Bielefeld, Germany}} \vspace*{0.4cm}
\protect\date \\ \vspace*{0.9cm}
{\bf   Abstract   \\ } \end{center} \indent
We investigate the three-dimensional $O(2)$ model near the critical point
by Monte Carlo simulations and calculate the major universal amplitude
ratios of the model. The ratio $U_0=A^+/A^-$ is determined directly from 
the specific heat data at zero magnetic field. The data do not, however,
allow to extract an accurate estimate for $\alpha$. Instead, we establish
a strong correlation of $U_0$ with the value of $\alpha$ used in the fit. 
This numerical $\alpha$-dependence is given by 
$A^+/A^- = 1 -4.20(5)\alpha+O(\alpha^2)$. For the special $\alpha$-values
used in other calculations we find full agreement with the corresponding
ratio values,  e.\ g.\ that of the  shuttle 
experiment with liquid helium. On the critical isochore we obtain the
ratio $\xi^+ /\xi^-_T=0.293(9)$, and on the 
critical line the ratio $\xi_T^c/\xi_L^c=1.957(10)$ for the amplitudes 
of the transverse and longitudinal correlation lengths. These two ratios
are independent of the used $\alpha$ or $\nu$-values. 

\vfill \begin{flushleft} 
Short title: Universal amplitude ratios of the $3d$ $O(2)$ model\\ 
\noindent{\rule[-.3cm]{5cm}{.02cm}} \\
\vspace*{0.2cm} 
E-mail: engels, holtmann, tschulze@physik.uni-bielefeld.de\,;\\
cucchieri, mendes@if.sc.usp.br
\end{flushleft} 
\end{titlepage}


\section{Introduction}

In quantum field theory and condensed matter physics $O(N)$ symmetric 
vector models play an essential part, because they are representatives of
universality classes for many physical systems. The universal properties 
of the $O(N)$ models - the critical exponents and amplitude ratios, which 
describe the critical phenomena -  are therefore of considerable importance.
In three dimensions the case $N=2$ is a special one: it is the first vector
model (with increasing $N$) showing Goldstone effects, and the exponent
$\alpha$, which controls the critical behaviour of the specific heat,
is very close to zero. In
fact, if one plots $\alpha$ versus $N$, as determined by field theory 
methods \cite{Zinn-Justin:2001bf}-\cite{Guida:1998}, then the function is 
approximately linear in $N$ and becomes negative just below $N=2$. The 
proximity of $\alpha$ to zero made it also difficult to determine the type 
of the singularity for the specific heat in real systems. Indeed, for the 
lambda transition of helium a nearly logarithmic singularity (corresponding
to $\alpha=0$) was first measured \cite{Fairbank:1957} and a similar 
behaviour was found at the gas-liquid critical point \cite{Bagatskii:1963}.
However, with the nowadays reached experimental precision, 
especially that of the spectacular shuttle experiment with
liquid helium \cite{Lipa:1996,Lipa:2000} there is no doubt that the 
critical exponent $\alpha$ is very small, but non-zero, and because it is
negative the peak of the specific heat is finite.

In this paper we calculate, among others quantities, the specific heat from 
Monte Carlo simulations. The determination of $\alpha$ from these data poses,
as we shall see, similar problems as in experiments. Of course, there is only
one value of $\alpha$ for the $3d$ $O(2)$-universality class, but it is 
unclear what the correct value is (see e.\ g.\ the survey in Table 19 of 
Ref.\ \cite{Pelissetto:2000ek}). We therefore pursue the strategy to 
calculate the universal ratios from our data for different $\alpha$-values
in the range where the actual value most probably is. The strongest
dependence on the used $\alpha$ is expected for fits involving the 
universal amplitude ratio $A^+/A^-$ of the specific heat. The same is
true for all theoretical 
determinations \cite{Campostrini:2001zw,Campostrini:2001iw} of this ratio.
Apart from $A^+/A^-$ we derive from our simulations other universal 
quantities and amplitude ratios, which characterize the $O(2)$-universality
class in three dimensions.

The model which we investigate is the standard $O(2)$-invariant 
nonlinear $\sigma$-model (or $XY$ model), which is defined by 
\be
\beta\,{\cal H}\;=\;-J \,\sum_{<x,y>} \phv_x\cdot \phv_y
         \;-\; {\vec H}\cdot\,\sum_{x} \phv_x \;.
\label{act}
\ee
Here $x$ and $y$ are the nearest-neighbour sites on a three-dimensional 
hypercubic lattice, $\phv_x$ is a $2$-component unit vector 
at site $x$ and $\vec H$ is the external magnetic field. We consider
the coupling constant $J$ as inverse temperature, that is $J=1/T$.
Instead of fixing the length of the spin vectors $\phv_x$ to 1 we could
have introduced an additional term $\sum_{x}[\phv_x^2 +\lambda (\phv_x^2
-1)^2]$ on the right hand side of the last equation. By choosing an
appropriate $\lambda$ value \cite{Hasenbusch:1999} it is then possible 
to eliminate leading order corrections to scaling. As it will turn out,
these corrections are negligibile in the energy density and 
marginal in the specific heat also with the Hamiltonian from Eq.
(\ref{act}). Moreover, we want to combine amplitudes obtained from 
former simulations at non-zero magnetic field \cite{Engels:2000xw} using
the same Hamiltonian with the amplitudes we determine now in order to 
calculate universal ratios.  

As long as $H=|{\vec H}|$ is non-zero one can decompose 
the spin vector $\phv_x$ into a longitudinal (parallel to the magnetic 
field ${\vec H}$) and a transverse component 
\be
\phv_x\; =\; \phi_x^{\parallel} {\vec{e}_H} + \phv_x^{\perp} ~,\quad
{\rm with}~~ {\vec{e}_H}= {\vec H}/H~. 
\ee
The order parameter of the system, the magnetization $M$, is then the 
expectation value of the lattice average $\phi^{\parallel}$
of the longitudinal spin component
\be
M \;=\; \langle\: \frac{1}{V}\sum_{x} \phi^{\parallel}_x\: \rangle\;
 =\; \langle \,  \phi^{\parallel}\, \rangle~.
\label{truem}
\ee
Here, $V=L^3$ and $L$ is the number of lattice points per direction.
There are two types of susceptibilities. The longitudinal 
susceptibility is defined as usual by the derivative of the magnetization, 
whereas the transverse susceptibility corresponds to the fluctuation 
of the lattice average $\phv^{\perp}$ of the transverse spin component
\ba
\chi_L\!\! &\!=\!&\!\! {\partial M \over \partial H}
 \;=\; V(\langle \: \phi^{\parallel 2}\: \rangle -M^2)~, \label{chil}\\
\chi_T\!\! &\!=\!&\!\! V \langle \: \phv^{\perp 2}\: \rangle 
~. \label{chit}
\ea
The total magnetic susceptibility is 
\be
\chi \;=\; \chi_L + \chi_T~.
\ee

At zero magnetic field, $H=0$, there is no longer a preferred direction and
the lattice average of the spins 
\be
\phv\;=\; \!\frac{1}{V}\sum_{x}\phv_x\;,
\ee
will have a vanishing expectation value on all finite lattices, 
$\langle \: \phv \: \rangle = 0$; the longitudinal and transverse
susceptibilities become equal for $T>T_c$ and diverge both for $T<T_c$
because of the Goldstone modes \cite{Engels:2000xw}. Nevertheless we 
can use $\phv$ to define the total susceptibility and the Binder 
cumulant by
\ba
\chi \!\!&\!=\!&\!\! V\langle \: \phv^2\: \rangle~,\label{chi}\\
g_r \!\!&\!=\!&\!\! {\langle \: (\phv^2)^2 \: \rangle \over
 \langle \: \phv^2 \: \rangle^2} -3~.
\label{gr}
\ea
For $T>T_c$ we have $\chi=2\chi_L=2\chi_T$. We approximate the order
parameter $M$ for $H=0$ by \cite{Talapov:1996yh} 
\be
M \;\simeq \; \langle |\phv|\, \rangle~.
\label{magmod}
\ee
On finite lattices the magnetization of Eq.\ (\ref{magmod})
approaches the infinite volume limit from above, whereas $M$
as defined by Eq.\ (\ref{truem}) for $H\ne 0$ reaches the 
thermodynamic limit from below.
 
In our zero field simulations we want to measure three further 
observables: the energy density, the specific heat and the 
correlation length. The energy of a spin configuration is simply
\be
E\;=\;-\sum_{<x,y>} \phv_x\cdot \phv_y\;,
\ee
and the energy density $\epsilon$ is then 
\be
\epsilon \;=\; \langle E \, \rangle / V \;.
\ee  
For the specific heat $C$ we obtain
\be
C \;=\; {\partial \epsilon \over \partial T}\;=\;
{J^2 \over V} \left( \langle E^2 \, \rangle -
\langle E \, \rangle^2 \right) \;.
\ee  
The second moment correlation length is calculated from the formula
\be
\xi_{2nd} \;=\; \left( {\chi / F -1 \over 4 \sin^2(\pi/L)} \right)^{1/2}~,
\label{correl}
\ee
where $F$ is the Fourier transform of the correlation function at
momentum $p_{\mu}=2\pi {\hat e}_{\mu}/L$, and ${\hat e}_{\mu}$ a unit
vector in one of the three directions
\be
F \;=\; {1 \over V} \langle | \sum_{x}\exp (ip_{\mu}x) \phv_x |^2
\rangle~.
\label{eff}
\ee
In the simulations we compute $F$ from an average over all three directions.
Strictly speaking, Eq.\ (\ref{correl}) can only serve as a definition 
of the correlation length for $T>T_c$, because the exponential correlation
length diverges for $H\rightarrow 0$ and $T<T_c$. Instead it is possible 
to introduce a transverse correlation length $\xi_T$ on the coexistence
line \cite{Privman:1991}, which is connected to the so-called stiffness
constant $\rho_s$ for $d=3$ by
\be
\xi_T  \;=\; \rho_s^{-1} \quad {\rm for}~ H=0,~~ T<T_c~. 
\ee
We explain later how to calculate $\rho_s$. For $H\ne 0$ there are
two exponential correlation lengths, a transverse ($\xi_T$) and a 
longitudinal one ($\xi_L$). Their second moment forms may be computed
 again from Eq. (\ref{correl}) by replacing $\chi$ and $F$ with their 
respective transverse or longitudinal counterparts.

The rest of the paper is organized as follows. First we discuss the
critical behaviour of the observables and define the universal 
amplitude ratios, which we want to determine. In Section 3 we 
describe our simulations at $H=0$, the results for the Binder cumulant,
the critical point and the correlation length. Then we analyse the data
for the energy and the specific heat. 
In Section 4 we discuss as an alternative the calculation of
$A^+/A^-$ from the equation of state, which was obtained from non-zero
field simulations. The following Section 5 serves to find the 
specific heat and the correlation lengths at $T_c$, as well as the 
stiffness constant, from $H\ne 0$ simulations. We close with a summary
of the ratios and the conclusions.


\section{Critical Behaviour}
\label{section:Criti}


In the thermodynamic limit ($V\rightarrow \infty$) the observables show
power law behaviour close to $T_c$. It is described by critical amplitudes
and exponents of the reduced temperature $t=(T-T_c)/T_c$. We note that we 
use here another definition of $t$ than in Ref.\ \cite{Engels:2000xw}.
We will mention this point again later. The scaling laws at $H=0$
are for:

\n the magnetization 
\be
 M  \;=\; B (-t)^{\beta} \quad {\rm for~} t<0~,
\ee
the longitudinal susceptibility
\be
 \chi_L \;=\; C^+ t^{-\gamma} \quad {\rm for~} t>0~,
\ee 
the transverse correlation length
\be
 \xi_T \;=\; \xi^-_T (-t)^{-\nu} \quad {\rm for~} t<0~,
\ee 
the correlation length
\be
 \xi \;=\; \xi^+ t^{-\nu} \quad {\rm for~} t>0~,
\label{xicr}
\ee
for $t \rightarrow \pm 0$ the energy density
\be
\epsilon \;=\; \epsilon_{ns} + T_c t \left (C_{ns}+
{A^{\pm} \over \alpha (1- \alpha)} |t|^{-\alpha} \right )~,
\ee 
and the specific heat
\be
 C \;=\; C_{ns} + {A^{\pm} \over \alpha} |t|^{-\alpha}~.
\label{spech}
\ee 
The specific heat and the energy density contain non-singular terms
$C_{ns}$ and $\epsilon_{ns}$, which are due to derivatives of the 
analytic part $f_{ns}$ of the free energy density. They are the
values of the specific heat and energy density at $T_c$. With our
definition for the specific heat amplitudes we have already singled
out their main $\alpha$-dependencies, the remaining factors $A^{\pm}$
are only moderately varying with $\alpha$.

On the critical line $T=T_c$ or $t=0$ we have for $H>0$ 
the scaling laws
\be
M \;=\; d_cH^{1/\delta} \quad {\rm or}\quad H \;=\;D_c M^{\delta}~,
\ee
and for the longitudinal and transverse correlation lengths
$\xi_{L,T}$
\be
\xi_{L,T} \;=\; \xi_{L,T}^c H^{-\nu_c}~,\quad
\nu_c\;=\; \nu /\beta\delta~.
\ee
The specific heat scales as
\be
C \;=\; C_{ns} + {A_c \over \alpha_c} H^{-\alpha_c}~, \quad 
\alpha_c\;=\; \alpha /\beta\delta~.
\ee 
We assume the following hyperscaling relations among the
critical exponents to be valid
\be
2-\alpha  \;=\; d\nu, \quad \gamma \;=\; \beta (\delta -1), \quad
d\nu \;=\; \beta (1 +\delta)~.
\ee
As a consequence only two critical exponents are independent. Because of
the hyperscaling relations and the already implicitly assumed equality
of the critical exponents above and below $T_c$ one can construct a
multitude of universal amplitude ratios \cite{Privman:1991}
(see also the discussion in Ref.\ \cite{Pelissetto:2000ek}). 
\n The following list of ratios contains those which we want to determine
here
\ba
&\!\!\!\! \!\! U_0  =\; A^+/A^-~, \quad &U_{\xi} 
\; = \; \xi^+ /\xi^-_T ~, \label{uratios}\\
&R_{\xi}^+ =\; (A^+)^{1/d} \xi^+ ~, \quad &R_{\xi}^T 
 =\; (A^-)^{1/d} \xi^-_T~,\label{rxiratios} \\
&\; R_{\chi} \; =\; C^+ D_c B^{\delta-1}~, \quad &R_C
\; =\; A^+ C^+ /B^2 ~,\label{rce}
\ea
and 
\be
R_A\; =\; A_c D_c^{-(1+\alpha_c)} B^{-2/\beta}~,\quad 
Q_2^T \; =\; (\xi_T^c /\xi^+)^{\gamma/\nu} C^+/ d_c(1/\delta +1)~.
\label{hratios}
\ee 
One of the ratios, $R_{\chi}$, was already calculated by us from non-zero
magnetic field simulations \cite{Engels:2000xw}, using the exponents
of Ref. \cite{Hasenbusch:1999}. We found
\be
R_{\chi}\; =\; 1.356(4)~.
\label{rchi}
\ee
In order to normalize the equation of state, the temperature and the 
magnetic field in the same paper, we had computed the critical amplitudes
of the magnetization on the coexistence line and the critical line 
with the result
\be
B\; =\; {\hat B} T_c^{\beta}\; =\; 1.245(7)~; \quad 
d_c\; =\; 0.978(2)~, D_c\; =\;1.11(1)~;
\label{camp}
\ee
where $ {\hat B}=0.945(5)$. The value for $J_c=T_c^{-1}=0.454165$ was taken
from Ref.\ \cite{Ballesteros:1996bd}.

\section{Simulations at $H=0$}
\label{section:FSSF}

\n All our simulations were done on three-dimensional lattices with periodic
boundary conditions. As in Ref.\ \cite{Engels:2000xw} 
we have used the Wolff single cluster algorithm. The main part of the $H=0$ data
was taken from lattices with linear extensions $L=24,36,48,72,96$ and 120. 
Between the measurements we performed 300-800 cluster updates to reduce the 
integrated autocorrelation time $\tau_{int}$. Apart from the largest lattice
($L=120$) where we made runs only at six couplings, we have generally scanned
the neighbourhood of $J_c$ by runs at more than 30 points on each lattice,
with special emphasis on the region $0.45414 \le J \le 0.45419$.
This enabled a comfortable reweighting analysis of the data. More details of
these simulations are presented in Table \ref{tab:survey}.  
\begin{table}
\begin{center}
\begin{tabular}{|r|c|c|c|c|c|c|}
\hline
$L$ & $J$-range & $N_J$ & $N_{meas} [1000]$   
& \multicolumn{1}{|c|}{$\tau_{int}( t<0)$ }
& \multicolumn{1}{|c|}{$\tau_{int}( t\approx 0)$ }
& \multicolumn{1}{|c|}{$\tau_{int}( t>0)$} \\
\hline
   24 & 0.440-0.4675 & 35 &  $\ge$ 100   & 1-3  & 1-3 & 1-3   \\
   36 & 0.440-0.4650 & 43 &  $\ge$ 100   & 1-4  & 2-3 & 2-10  \\
   48 & 0.442-0.4650 & 55 &  $\ge$ 100   & 1-5  & 2-5 & 4-13  \\
   72 & 0.4465-0.460 & 41 &   80-100     & 1-4  & 4-8 & 7-21  \\
   96 & 0.450-0.4567 & 33 &   60-80      & 2-10 & 6-7 & 7-35  \\
  120 & 0.452-0.4562 &  6 &    20        & 2-4  & 14  & 12-23 \\
\hline
\end{tabular}
\end{center}
\caption{Survey of the Monte Carlo simulations  at $H=0$ for different
lattices. Here $N_J$ is the number of different couplings at which
runs were performed; $\tau_{int}$ is the integrated autocorrelation
time for the energy and $N_{meas}$ the number of measurements per 
coupling in units of 1000.}
\label{tab:survey}
\end{table}

\subsection{The Critical Point and the Binder Cumulant}

It is obvious that any determination of critical amplitudes relies crucially
on the exact location of the critical point. Since we have produced a 
considerable amount  

\begin{figure}[b!]
\begin{center}
   \epsfig{bbllx=63,bblly=265,bburx=516,bbury=588,
       file=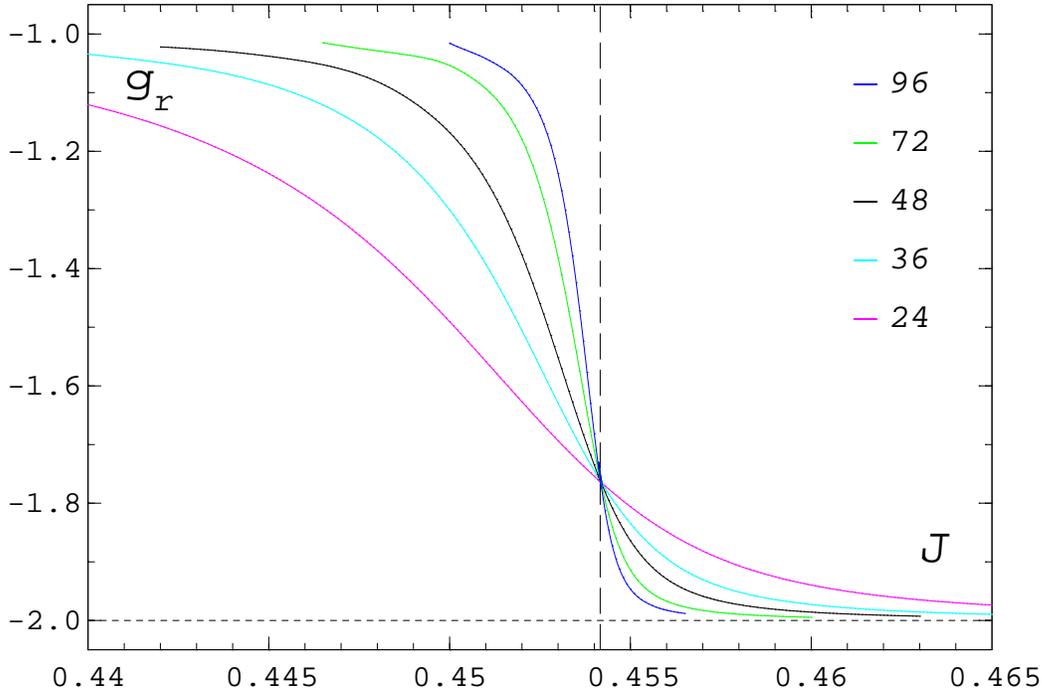, width=120mm}
\end{center}
\caption{The Binder cumulant $g_r$ from Eq.\ (\ref{gr}) as a function of
the coupling $J$. The curves were obtained by reweighting the data.
With increasing lattice size $L= 24,36,48,72$ and 96, the slope of the 
respective curve increases close to the critical point. The vertical 
dashed line denotes $J_c$ of Ref.\ \cite{Ballesteros:1996bd}.}
\label{fig:gr}
\end{figure}
\newpage
\begin{figure}[t!]
\begin{center}
   \epsfig{bbllx=63,bblly=265,bburx=516,bbury=588,
       file=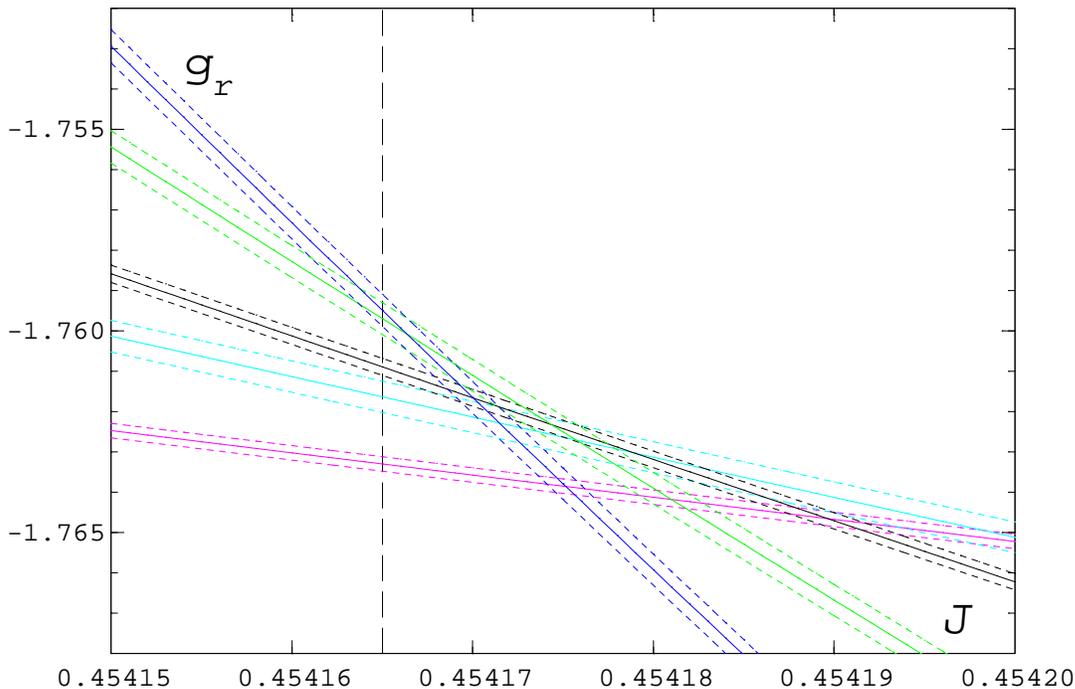, width=120mm}
\end{center}
\caption{The Binder cumulant $g_r$ in the close neighbourhood of the
critical point. The figure is an enlargement of Fig.\ \ref{fig:gr}.
The dashed lines accompanying the solid lines show the jackknife error 
corridor.}
\label{fig:grbig}
\end{figure}

\begin{figure}[t!]
\begin{center}
   \epsfig{bbllx=63,bblly=265,bburx=516,bbury=588,
       file=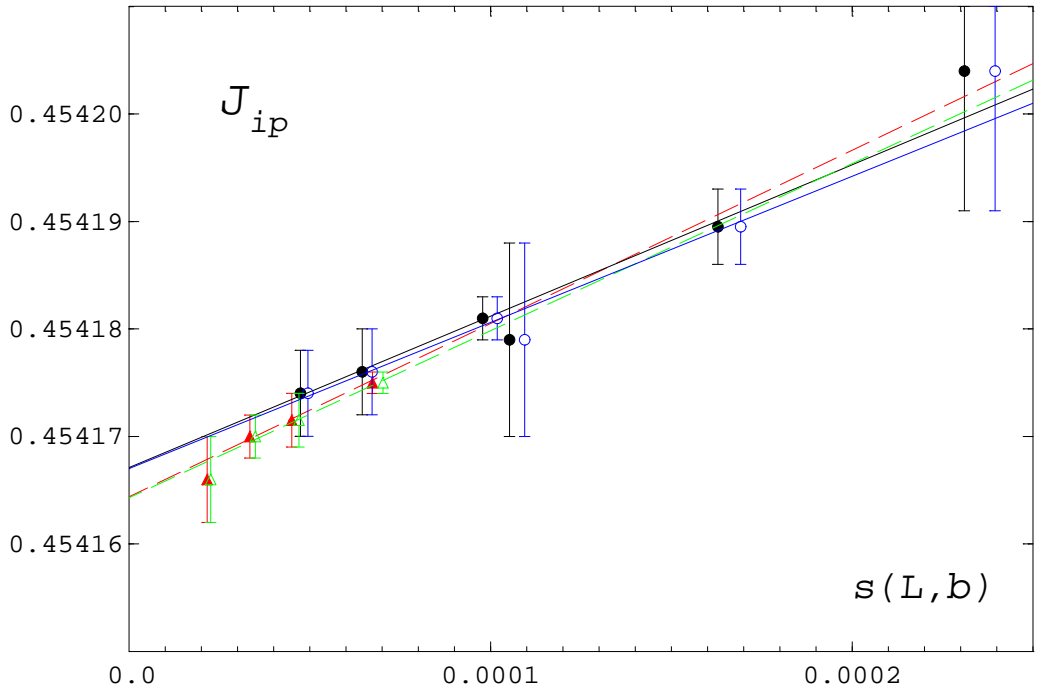, width=120mm}
\end{center}
\caption{The coupling $J_{ip}$ at the intersection point of $g_r(L)$ and 
$g_r(bL)$ for various combinations of $L$ and $b$ as a function of 
$s(L,b)$, Eq.\ (\ref{shift}). The filled (empty) symbols were calculated 
with $\nu=0.669~(0.673)$. The dashed (solid) lines are linear fits with 
(without) the $L=96$ intersection points, denoted here by triangles.}
\label{fig:jc}
\end{figure}

\n of data in the neighbourhood of the critical point it 
was natural to verify first the rather precise result of Ballesteros et al.\
\cite{Ballesteros:1996bd}. We have done this by studying the Binder
cumulant $g_r$, which is directly a finite-size-scaling function
\be
g_r \;=\; Q_g (t L^{1/\nu}, L^{-\omega})~.
\label{grscl} 
\ee
The function $Q_g$ depends on the thermal scaling field and on possible
irrelevant scaling fields. Here we have specified only the leading irrelevant 
scaling field proportional to $L^{-\omega}$, with $\omega>0$.  
At the critical point, $t=0$, $g_r$ should therefore be independent of $L$
apart from corrections due to these irrelevant scaling fields. In Fig.\
\ref{fig:gr} we show our results for $g_r$ as obtained by reweighting the
direct data. We observe, at least on the scale of Fig.\ \ref{fig:gr}, no
deviation from the scaling hypothesis. However, after a blow-up of the 
close vicinity of the critical point, as shown in Fig.\ \ref{fig:grbig},
we can see that the intersection points between curves from different 
lattices are not coinciding. The shift $\Delta J$ of the crossing point
from the infinite volume critical coupling $J_c$ can be estimated
by expanding the scaling function $Q_g$ to lowest order in both variables.
For two lattices with sizes $L$ and $L\p=bL$ one gets
\be
\Delta J^{L,L\p} \;\propto \; s(L,b)\; =\; { 1 -b^{-\omega} \over 
b^{1/\nu} -1} L^{-\omega -1/\nu}~.
\label{shift}
\ee
In Fig.\ \ref{fig:jc} we
have plotted the $J$-values of the intersection points for each pair of
lattices as a function of the variable $s(L,b)$ of Eq.\ (\ref{shift}).
For $\omega$ we used the value 0.79(2) of Ref.\ \cite{Hasenbusch:1999},
and for $\nu$ we have chosen the two values $\nu=0.669$ and $0.673\:$ as
bounds of the probable $\nu$-range. Of course, the intersection points are
completely independent of $\nu$ and $\omega$. Only the variable $s(L,b)$
is changing when the exponents are changed. As can be seen in Fig.\ 
\ref{fig:jc} also the extrapolation to the critical point $J_c$ for 
$L\rightarrow 0$ (or $s(L,b)\rightarrow 0$) is unaffected by the 
choice of $\nu$. The same applies to a variation of $\omega$.
Since the slope of $g_r(L=96)$ close to the critical
point is rather large, a small numerical uncertainty might shift the 
intersection points with the other curves considerably. We have therefore
determined $J_c$ also by fits excluding the results from the largest
lattice. Thus we arrive at the final estimate
\be
J_c\;= \; 0.454167(4),
\ee
in full agreement with the result $J_c=0.454165(4)$ of Ballesteros
et al.\ \cite{Ballesteros:1996bd}. In order to be consistent with our 
previous papers we use in the following again the value of Ref.\
\cite{Ballesteros:1996bd}.

In a similar manner one can determine from the same data the universal
value $g_r(J_c)$. The difference of the $g_r$-values at the intersection
points to $g_r(J_c)$ is here
\be
\Delta g_r^{L,L\p} \;\propto \; s_o(L,b)\; =\; {b^{1/\nu} -b^{-\omega} \over 
b^{1/\nu} -1} L^{-\omega}~.
\label{shifto}
\ee
\newpage
\begin{figure}[t!]
\begin{center}
   \epsfig{bbllx=63,bblly=265,bburx=516,bbury=588,
       file=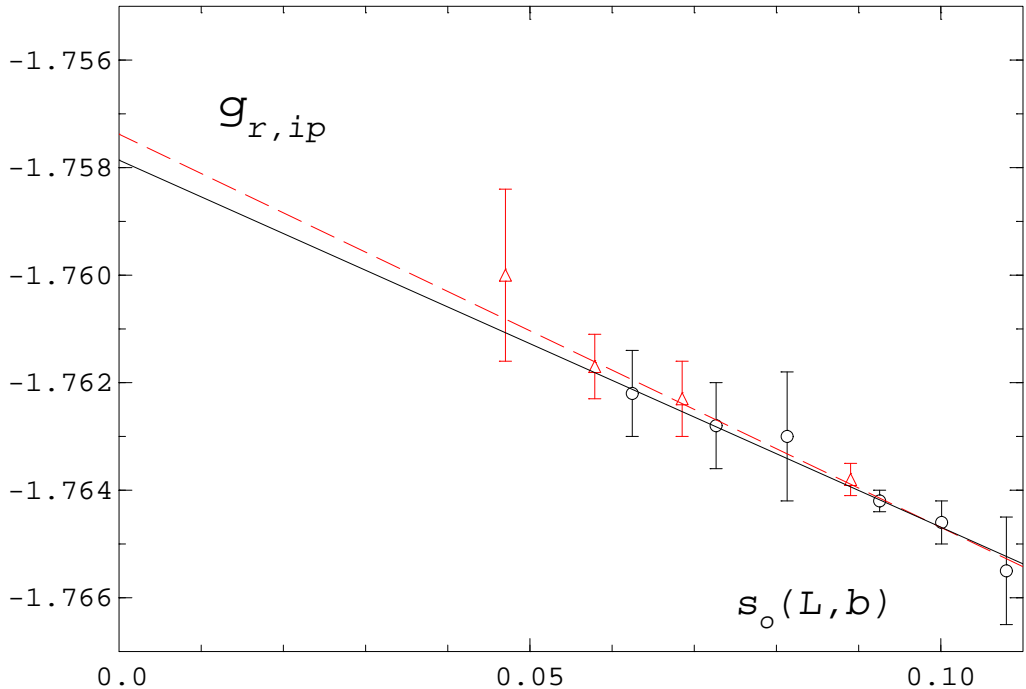, width=120mm}
\end{center}
\caption{The Binder cumulant $g_{r,ip}$ at the intersection point
for various combinations of $L$ and $b$ as a function of $s_o(L,b)$, 
Eq.\ (\ref{shifto}). The dashed (solid) lines are linear fits with 
(without) the $L=96$ intersection points, denoted here by triangles.}
\label{fig:grjc}
\end{figure}
\n In Fig.\ \ref{fig:grjc} we show the extrapolation of $g_r$ to the 
critical point value at $s_o(L,b)=0$. A variation of $\omega$ in the 
range 0.77-0.81 leads only to a shift of $10^{-4}$. The new variable 
$s_o(L,b)$ is practically independent of $\nu$, the influence of $\nu$ 
is not visible in Fig.\ \ref{fig:grjc}. Comparing again extrapolations
with and without the $L=96$ points one obtains 
\be
g_r(J_c)\; = \; -1.758(2)~,\quad {\rm or}\quad  {\langle \: (\phv^2)^2 
\, \rangle \over \langle \: \phv^2 \: \rangle^2}(J_c)
\; =\; 1.242(2)~, 
\label{gratjc}
\ee 
well in accord with the result of Ref.\ \cite{Campostrini:2001iw}
(see also the long discussion in Ref.\ \cite{Arnold:2001ir}).

\subsection{The Correlation Length}

\begin{figure}[t!]
\begin{center}
   \epsfig{bbllx=63,bblly=265,bburx=516,bbury=588,
       file=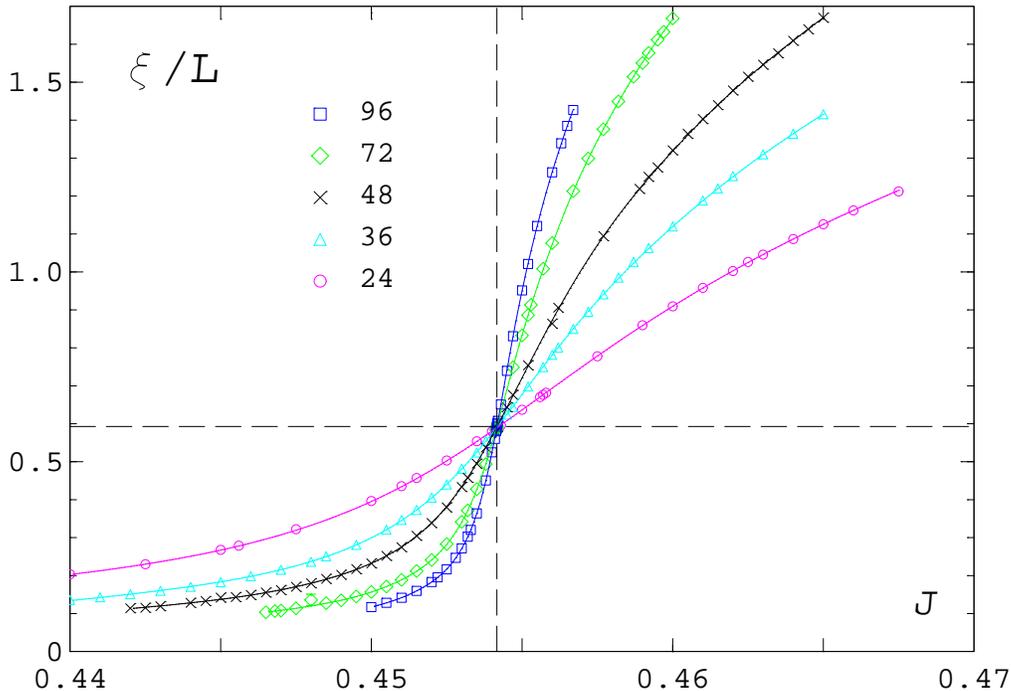, width=120mm}
\end{center}
\caption{The correlation length $\xi$ divided by $L$ versus $J$
for $L=24,36,48,72$ and 96. The solid lines were calculated by
reweighting the data. The dashed vertical line gives the postion of
$J_c$, the horizontal one the universal value, Eq. (\ref{xiatjc}).}
\label{fig:xiol}
\end{figure}

In our $H=0$ simulations we have measured the correlation length using
the second moment formula, Eq. (\ref{correl}). The finite-size-scaling
equation for $\xi$ is
\be
\xi \;=\; L Q_{\xi} (t L^{1/\nu}, L^{-\omega})~,
\label{xiscl} 
\ee
and $\xi /L=Q_{\xi}$ is a scaling function like $g_r$, that is its value 
at the critical point is universal for $L\rightarrow \infty$. In Fig.\
\ref{fig:xiol} we have plotted our correlation length data divided by $L$.
Here formula (\ref{correl}) has also been evaluated for $J>J_c$ or $T<T_c$
though in this region the data cannot be identified with the correlation
length. We see again that all curves intersect at the previously determined
critical point. A closer look into the neighbourhood of $J_c$ reveals 
however similar corrections to scaling as in the case of $g_r$. The 
corresponding extrapolation of the variable $s_o(L,b)$ to zero leads for 
$\xi/L$ to 
\be
\xi / L \:(J_c)\;=\; 0.593(2)~.
\label{xiatjc}
\ee
This result confirms nicely the value  $\xi/L = 0.5927$ from the preliminary
simulations mentioned in Ref.\ \cite{Hasenbusch:1999}.   

Our data for the correlation length can also be used to find the critical
amplitude $\xi^+$ of Eq. (\ref{xicr}). To this end we use a method
described in detail in Ref.\ \cite{Engels:1999nv}. We briefly repeat the
main arguments assuming for simplicity that there are no corrections to scaling.
An observable $O$ with critical behaviour approaches for either positive or
negative $t$ and $L\rightarrow \infty$ the limiting form
\be
O_{\infty} \;= \; a_0 |t|^{-\rho}~,\quad {\rm for~} |t| \rightarrow 0~,
\label{oinf}  
\ee
where $a_0$ is the critical amplitude and $\rho$ the critical exponent. At
finite $L$ the observable satisfies a scaling relation
\be
O(t,L) \;= \; L^{\rho/\nu} Q_O(x_t)~,\quad {\rm with~} x_t\,=\, tL^{1/\nu}~. 
\label{ofin}
\ee
Here, $Q_O$ is the finite-size-scaling function of $O$. In order to ensure
the correct thermodynamic limit for fixed small $|t|$ we must have the 
relation
\be
O_{\infty} \;= \;|t|^{-\rho} \lim_{x_t\to\pm\infty} |x_t|^{\rho}Q_O(x_t)~.
\label{oho}
\ee
The sign of $x_t$ is of course the same as that of $t$. It is clear then, that
the function 
\be
A_O(x_t) \;=\; |x_t|^{\rho}Q_O(x_t)~,
\label{afun}
\ee
will converge asymptotically to the critical amplitude $a_0$. Moreover,
$a_0$ will be an extreme value of $A_O(x_t)$.

\begin{figure}[b!]
\begin{center}
   \epsfig{bbllx=63,bblly=280,bburx=516,bbury=563,
       file=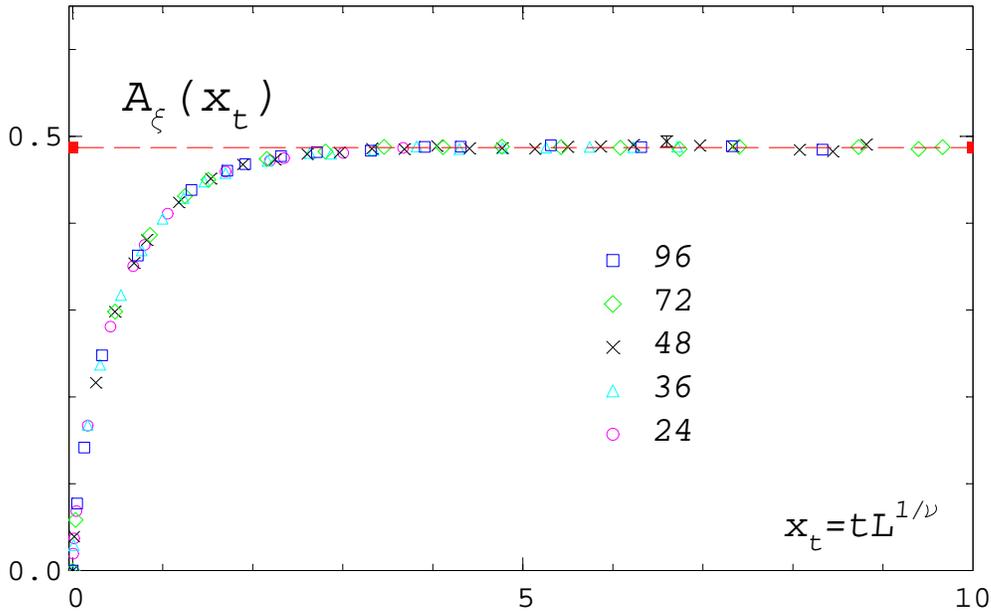, width=120mm}
\end{center}
\caption{The amplitude function $A_{\xi}$, Eq.\ (\ref{afun}), of the
correlation length versus the scaling variable $x_t$ for
$\nu=0.671$ and $L=24,36,48,72$ and 96.
The horizontal line indicates the $\xi^+$-value.}
\label{fig:axi}
\end{figure}

\begin{figure}[t!]
\begin{center}
   \epsfig{bbllx=63,bblly=280,bburx=516,bbury=563,
       file=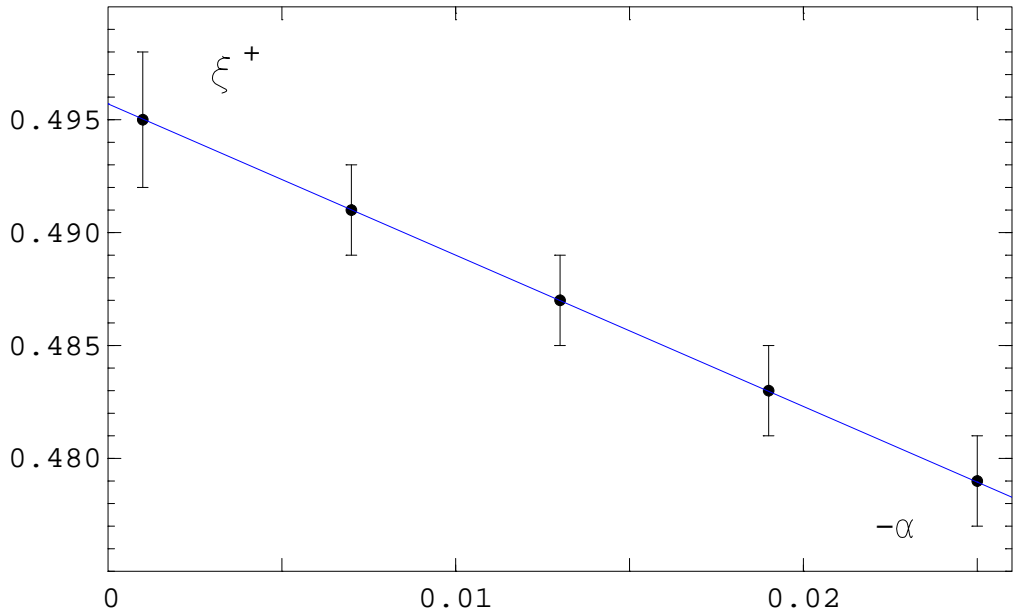, width=120mm}
\end{center}
\caption{The critical amplitude $\xi^+$, Eq.\ (\ref{xicr}), of the
correlation length versus $-\alpha$. The data (circles) are determined
from the amplitude function $A_{\xi}(x_t)$, the solid line is the linear
fit (\ref{xiplus}).}
\label{fig:xip}
\end{figure}

We have applied this method to the correlation length results.
In Fig.\ \ref{fig:axi} we show $A_{\xi}(x_t)$ for the exponent $\nu=0.671$
and various $L$-values. We notice that already at $x_t\approx 4$ a plateau
is reached and essentially no corrections to scaling are visible. The 
marginal spread of the data in the plateau region leads only to a small
error for the amplitude $\xi^+$. Since the scaling variable $x_t$ changes
with $\nu$ there is however a $\nu$-dependence, which can also be expressed
as a dependence on $\alpha$, because of the hyperscaling relation
$2-\alpha=d\nu$. In fact, after evaluating $A_{\xi}$ for several 
$\nu$-values, we find that $\xi^+$ is rather exactly a linear function
of the used $\alpha$
\be
\xi^+ \;=\; 0.4957(20) +0.67(12)~\alpha~.
\label{xiplus}
\ee
This can be seen in Fig.\ \ref{fig:xip}, where we compare the fit, Eq.\
\ref{xiplus}, to some directly determined $\xi^+$-values.

\subsection{Specific Heat and Energy Density at $T_c$}
\label{section:sphattc}

\begin{figure}[ht]
\begin{center}
   \epsfig{bbllx=63,bblly=280,bburx=516,bbury=563,
       file=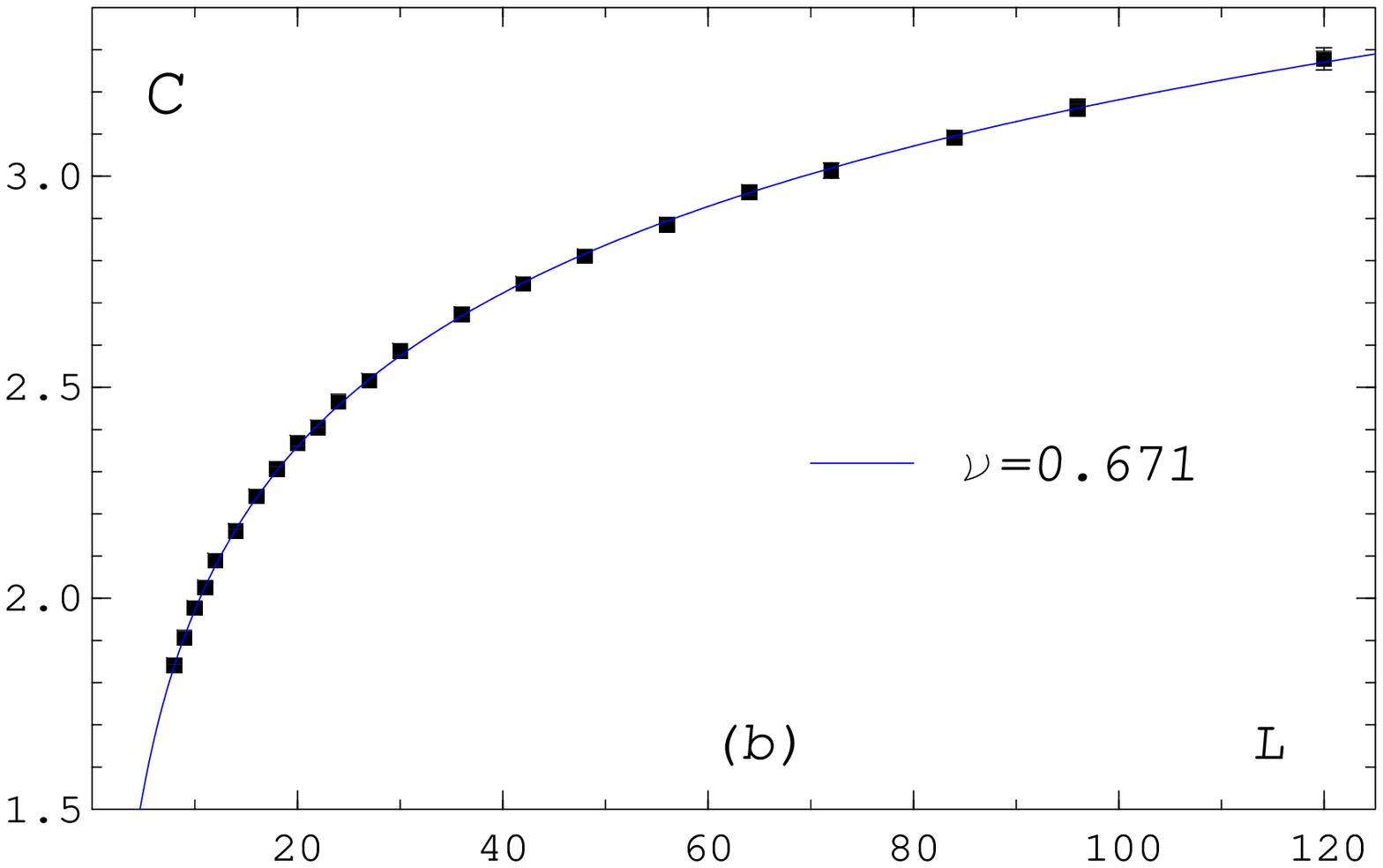, width=120mm}
\end{center}
\caption{The energy density (a) and the specific heat (b) versus $L$
at the critical point. The dashed line shows $\epsilon_{ns}$ and the
solid lines fits to Eqs. (\ref{epsjc}) and (\ref{cjc}) for $\nu=0.671$
and $\omega=0.79$.}
\label{fig:eandc}
\end{figure}

As mentioned already in Section \ref{section:Criti} both the energy density
and the specific heat contain additional non-singular terms. This fact
complicates of course the determination of the critical amplitudes. We can 
however calculate the non-singular terms beforehand by a finite-size-scaling
analysis directly at the critical point. For that purpose we have made 
further Monte Carlo runs at $T_c$ on 23 lattices with $L=8$ to $L=160$. 
In these runs we took between 500,000 and $200,000$ measurements each
for $L=8-64$ and on the larger lattices between 120,000 and 50,000. The data
for the energy density and the specific heat are shown in Fig.\ \ref{fig:eandc}
as a function of $L$ up to $L=120$. If one expands the scaling functions 
for $\epsilon$ and $C$ at $T_c$ in powers of $L^{-\omega}$ one obtains 
\ba
\epsilon (L) &\!\! = \!\!& \epsilon_{ns}\: + q_{0\epsilon} L^{(\alpha -1)/\nu} 
\left ( 1+  q_{1\epsilon} L^{-\omega} + \dots \right)~, 
\label{epsjc} \\
C(L) &\!\! =\!\! & C_{ns} + q_{0C} L^{\alpha /\nu} 
\left ( 1+  q_{1C} L^{-\omega} + \dots \right)~.
\label{cjc}
\ea
We have fitted the first terms (up to $q_1$) of these expansions to the data.
In the case of the energy density we find no corrections to scaling, that is
$q_{1\epsilon} \approx 0$, and only small corrections for the specific heat.
Fits with different $\nu$-values cannot be distinguished in Fig.\ 
\ref{fig:eandc}.
When we treat $\nu$ as a free fit parameter we get $\nu=0.671(2)$. 
The quantity $\epsilon_{ns}$ exhibits no noticeable dependency on $\nu$ or 
$\alpha$ and $\omega$. We find
\be
\epsilon_{ns}\; = \; -0.98841(3)~.
\label{ens}
\ee
\begin{figure}[t!]
\begin{center}
   \epsfig{bbllx=63,bblly=280,bburx=516,bbury=563,
       file=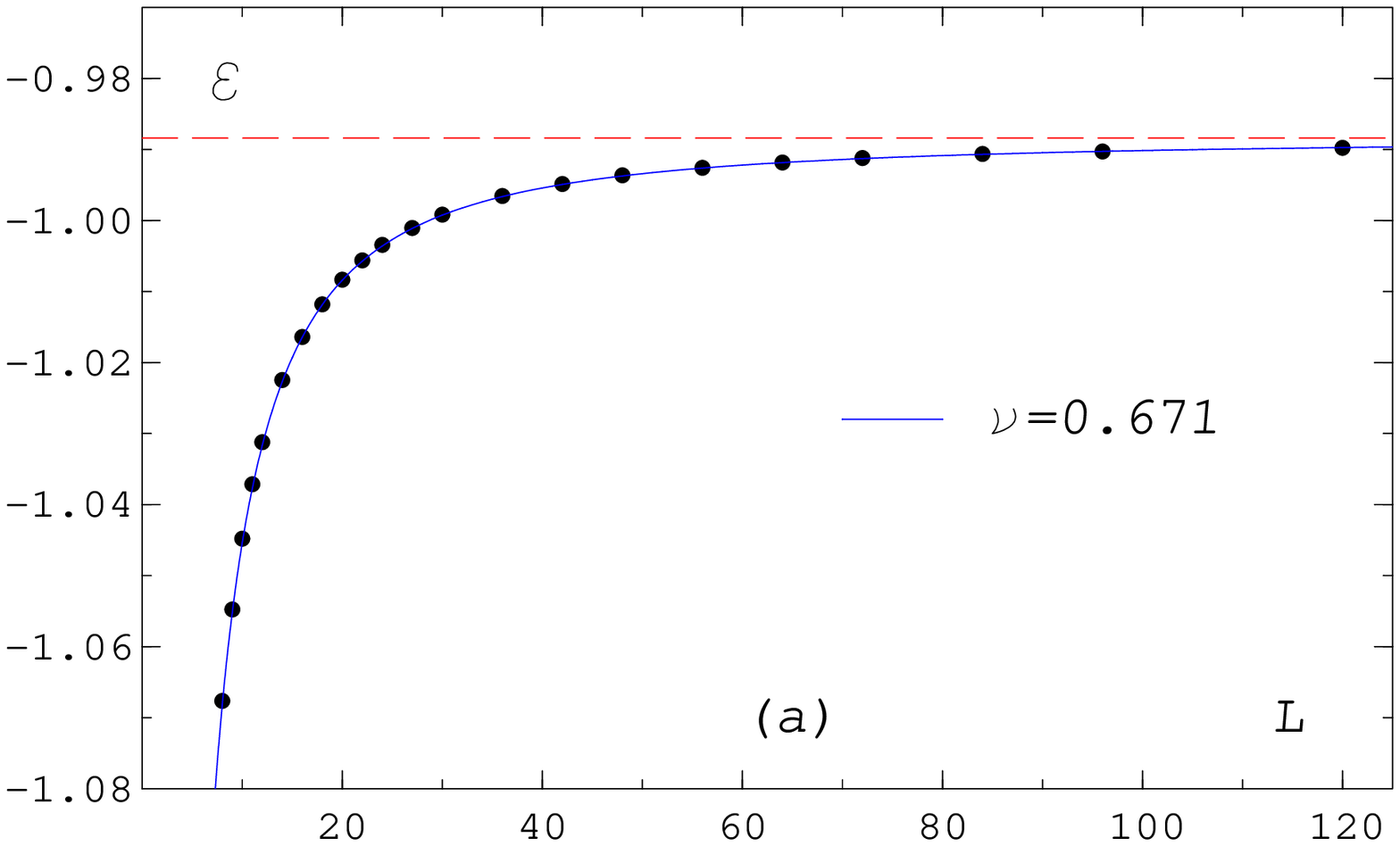, width=120mm}
\end{center}
\end{figure}

\vspace{-0.3cm}
\n The situation is quite different in the case of the specific heat. Its 
non-singular part varies from about 50 for $\nu=0.669$ to 16 at $\nu=0.675$. 
The reason for this strong variation is that the exponent $\alpha
=2-3\nu$ is close to zero, when $\nu$ approaches 2/3. Then the background
term $C_{ns}$ develops a pole ($\sim 1/\alpha$) which cancels a corresponding
pole in the critical amplitude in such a way that the characteristic critical
power behaviour ($\sim |t|^{-\alpha}$) turns over into a logarithmic behaviour
($\sim \ln |t|$).
\begin{figure}[t!]
\begin{center}
   \epsfig{bbllx=63,bblly=280,bburx=516,bbury=563,
       file=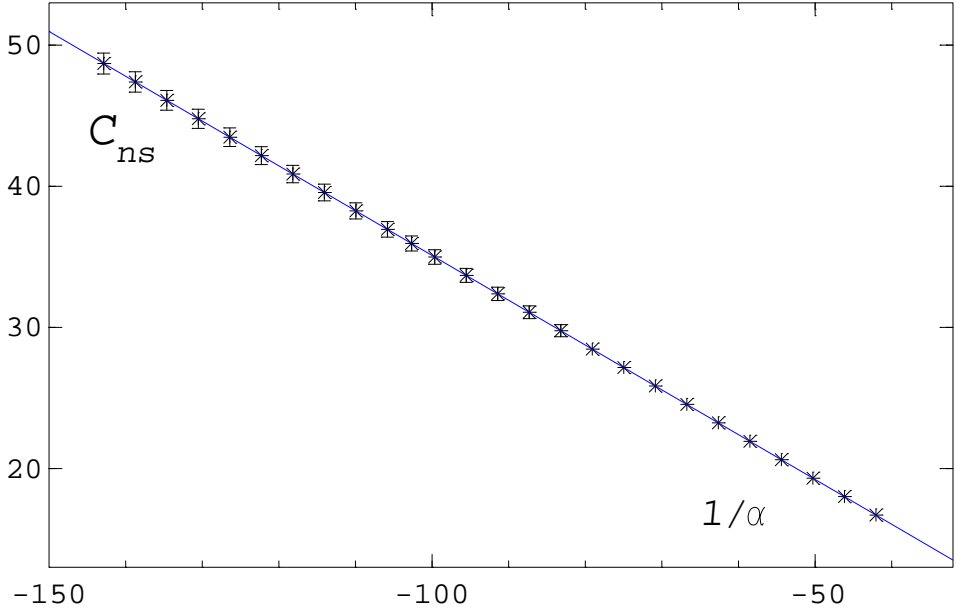, width=120mm}
\end{center}
\caption{The non-singular part $C_{ns}$ of the specific heat versus
$1/\alpha$ from fits to Eq.\ (\ref{cjc}) (stars) with $\omega=0.79$. 
The solid line is from Eq.\ (\ref{cnsfit}).}
\label{fig:cns}
\end{figure}
This mechanism for the emergence of the logarithmic singularity as 
$\alpha \to 0$ is well-known (see Refs.\ \cite{Privman:1991} and
\cite{Widom:1965,Wegner:1972}). We demonstrate it by assuming that 
\ba
C_{ns}(\alpha) &\!\!=\!\!& c_{ns}^0 + { c_{ns}^{p} \over \alpha}~,
\label{cnsexp}\\
A^{\pm}(\alpha)&\!\!=\!\!& a_0^{\pm} + a_1^{\pm}\alpha + O(\alpha^2)~.
\label{Aexp}
\ea
If we insert these equations into Eq.\ (\ref{spech}) and expand 
$|t|^{-\alpha}$ for small $\alpha$ we obtain
\ba
C &\!\!=\!\!& c_{ns}^0 + { c_{ns}^{p} \over \alpha} +
\left( {a_0^{\pm} \over \alpha} + a_1^{\pm}  + O(\alpha) \right )
\left(1 -\alpha \ln |t| +\dots \right )
\label{Cexp}\\
  &\!\!=\!\!& c_{ns}^0 +{ c_{ns}^{p} + a_0^{\pm} \over \alpha}  + a_1^{\pm}
 -a_0^{\pm} \ln |t| +  O(\alpha)~.
\label{Cexp2}
\ea
Evidently the limit of $C$ for $\alpha \to 0$ exists and has a logarithmic 
$|t|$-dependence, if the pole term vanishes, which requires \cite{Widom:1965}
\be
c_{ns}^{p}\;= \; -a_0^{\pm}, \quad {\rm and}\quad  a_0^{+} \;= \; a_0^{-}~.
\label{polcon}
\ee
The ratio $A^+/A^-$ is therefore close to 1
\be
A^+/A^-\;=\; 1+O(\alpha)~.
\ee
In Fig.\ \ref{fig:cns} we show the non-singular part $C_{ns}$ of the specific 
heat resulting from fits to Eq.\ (\ref{cjc}) with $\omega=0.79$ and various
values for $\alpha$ plotted versus $1/\alpha$. The $\chi^2$ per degree of 
freedom in each fit is 0.83(1), preferring no particular $\alpha$-value. We
see that indeed $C_{ns}$ is linearly dependent on $1/\alpha$. A fit to the 
ansatz, Eq.\ (\ref{cnsexp}), gives
\be
C_{ns} \;=\; 3.35(4) - {0.3175(5) \over \alpha}~,
\label{cnsfit}
\ee
with an extremely small $\chi^2/N_f$ of the order of $10^{-4}$. We conclude
from this fact, that the pole term behaviour of $C_{ns}$ is not a numerical 
accident, but underlines the previous considerations. In order to study the
influence of the correction exponent $\omega$ we have repeated the whole
analysis of $C(L)$ for the values $\omega=0.77$ and $\omega=0.81$, that is
a standard deviation away from the central value 0.79. The $\chi^2/N_f$ for
each single fit to Eq.\ (\ref{cjc}) is again 0.83(1), the new values for
$C_{ns}$ coincide within error bars with the values for $\omega=0.79$,
however the resultant linear fits in $1/\alpha$ to Eq.\ (\ref{cnsexp})
at fixed $\omega$, lead to slight changes (again with
a $\chi^2/N_f$ of the order of $10^{-4}$)  
\be
C_{ns} = \Biggl\{\matrix { 3.37(4)-0.3165(5)/\alpha & {\rm for} &
 \omega=0.77 \cr
 3.33(4)-0.3184(5)/\alpha & {\rm for} &  \omega=0.81 \cr}~~,
\label{omvar}
\ee
mainly for the pole term parameter $c_{ns}^{p}$.

In the following we shall use the results for $C_{ns}$ to analyze as well
the specific heat data for $T\ne T_c$. If not explicitly mentioned, the
fit results have always been obtained for fixed $\omega=0.79$. We have 
repeated the following analysis also for $\omega=0.77$ and 0.81 and shall
comment on any noticeable changes due to $\omega$.
\begin{figure}[t!]
\begin{center}
   \epsfig{bbllx=103,bblly=263,bburx=508,bbury=588,
       file=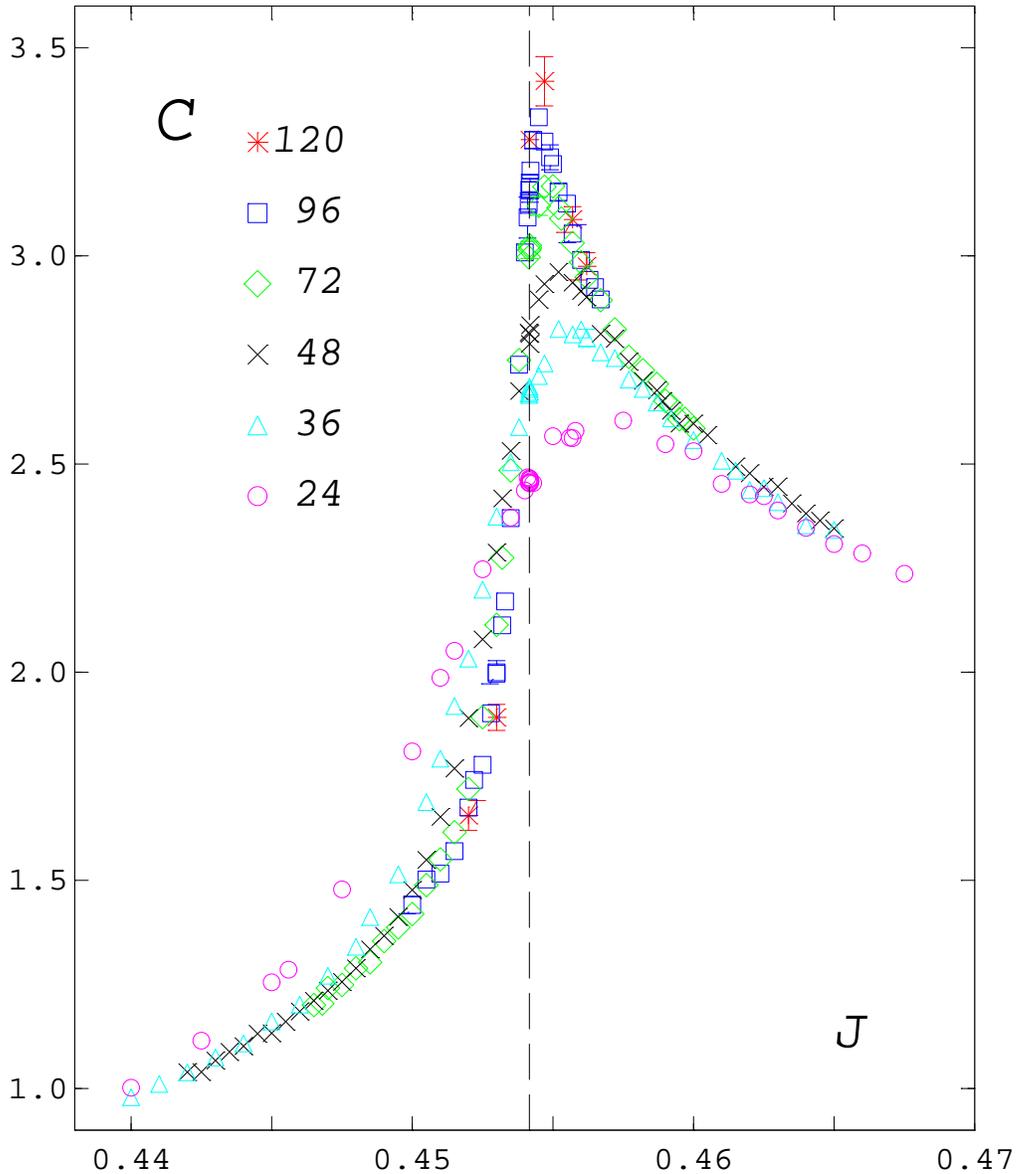,height=120mm,angle=-90}
\end{center}
\vspace{0.3cm}
\caption{The specific heat data for different $L$ versus the coupling $J$.
The dashed line indicates the position of the critical point.}
\label{fig:cv1}
\end{figure}

\subsection{The Specific Heat and $A^+/A^-$}
\label{section:heatratio}
\begin{figure}[ht]
\vspace*{0.2cm}
\begin{center}
   \epsfig{bbllx=63,bblly=280,bburx=516,bbury=563,
       file=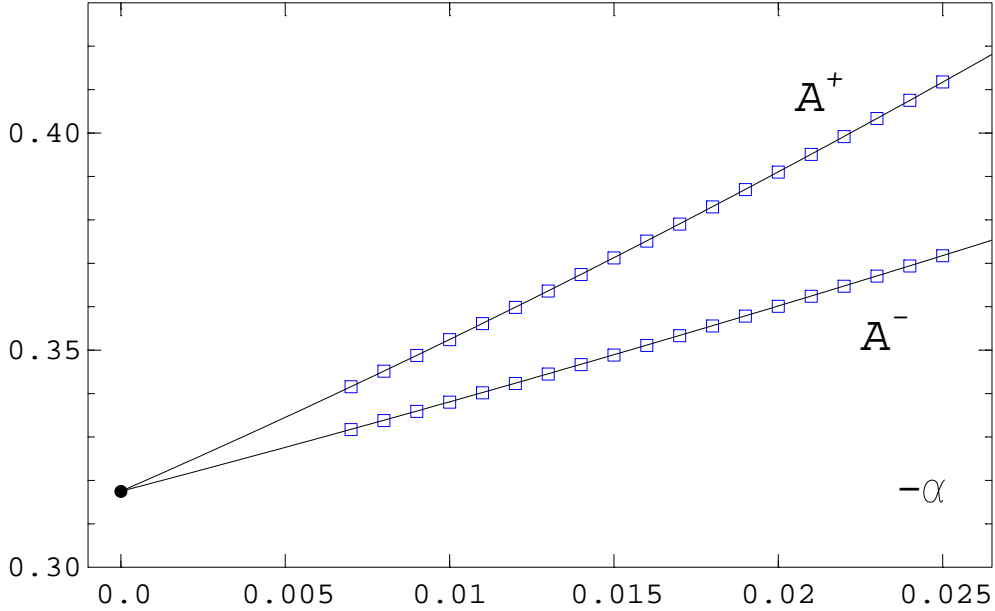, width=120mm}
\end{center}
\caption{The amplitudes $A^+$ and $A^-$ versus $-\alpha$ (squares). The
filled circle is the value expected from $C_{ns}$; the lines are the 
parametrizations (\ref{aplus}) and (\ref{aminus}).}
\label{fig:apam}
\end{figure}

In Fig.\ \ref{fig:cv1} we have collected all our specific heat data 
at zero magnetic field for the $L$-values of Table \ref{tab:survey}. We observe
with increasing $L$ a more and more pronounced peak close to $J_c$. As
already discussed in the introduction, we nevertheless expect a finite peak
height even in the thermodynamic limit, since the singular part of $C$
vanishes at the critical point for negative $\alpha$. The peak (and not dip)
behaviour implies also that the amplitude $A^{\pm}/\alpha$ must be negative,
or that $A^{\pm}$ is positive. The previous analysis of the non-singular 
contribution to $C$ confirms this consideration: because $c_{ns}^p$ is
negative we have a positive value $a_0^{\pm}=a_0$ for the leading part 
of $A^{\pm}$. 
\begin{figure}[t!]
\begin{center}
   \epsfig{bbllx=63,bblly=265,bburx=516,bbury=588,
       file=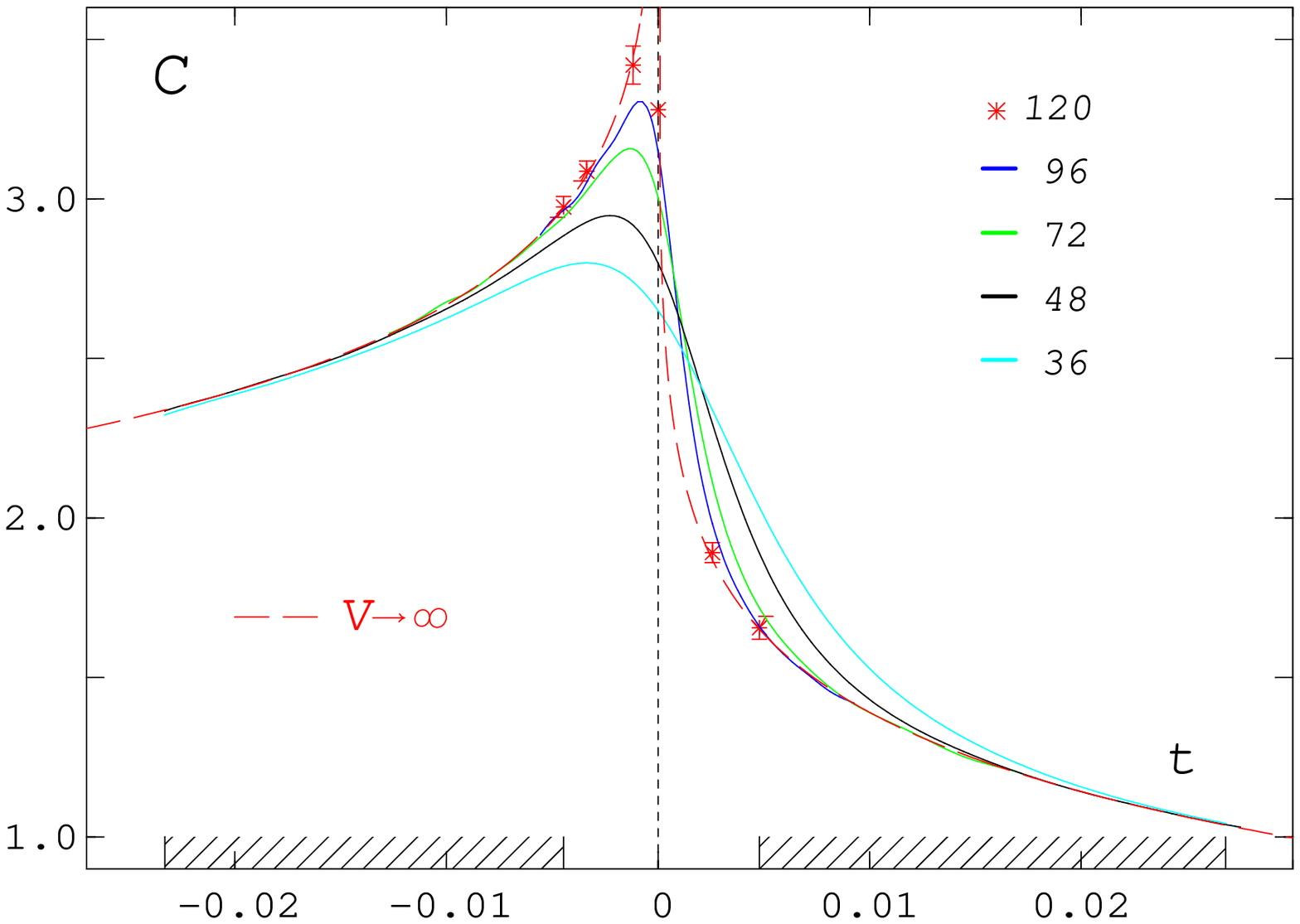, width=120mm}
\end{center}
\caption{The specific heat versus the reduced temperature $t$ for $L=36,$
48, 72, 96 and 120 (stars). The solid lines were calculated by
reweighting the data, the peak height increases with $L$. The line
of long dashes is the fit from the ansatz, Eq.\ (\ref{spfit}), for 
$\alpha=-0.013$ and $\omega=0.79$. The hatched areas show the fit regions.}
\label{fig:cv2}
\end{figure}
We have interpolated the data points by reweighting, apart from the $L=120$
results. The respective curves are plotted in Fig.\ \ref{fig:cv2} as a 
function of $t$. Compared to Fig.\ \ref{fig:cv1} we have therefore an
exchange of the high ($t>0,J<0)$ and low temperature ($t<0,J>0)$ parts in the 
figures. In order to find the amplitudes $A^{\pm}$ we have made the following
ansatz including correction-to-scaling terms
\be
 C \;=\; C_{ns} + {A^{\pm} \over \alpha} |t|^{-\alpha}
\left ( 1 + c_1^{\pm}|t|^{\omega\nu} + c_2^{\pm} t \right )~.
\label{spfit}
\ee 
For a fit to the form (\ref{spfit}) the curves from the largest 
lattices were used in those $t$-ranges, which appear hatched in Fig.\ 
\ref{fig:cv2}, that is for $-0.0233\le t\le -0.0045$ 
and $0.0048\le t\le 0.0268$. The non-singular part from Eq.\ (\ref{cnsfit})
was then taken as an input to the fit, whereas the $L=120$ data points
served only as a check of the fit result. As an example we
show in Fig.\ \ref{fig:cv2} the fit for $\alpha=-0.013$. Fits with other
small, negative $\alpha$-values work as well and have the same $\chi^2$
per degree of freedom, namely $1.03\;$. In Table \ref{tab:cfitr} we present
details of the fits for several $\alpha$-values. The two correction-to-scaling
contributions are always opposite in sign and cancel therefore to some extent,
especially in the high temperature region. The amplitudes $A^{\pm}$ are
still $\alpha$-dependent, though in our notation we have taken the 
anticipated pole behaviour already into account. We find that $A^+$ and  
$A^-$ are nearly linear functions of $\alpha$. The $\alpha$-dependence
of the fit results for the amplitudes is shown in Fig.\ \ref{fig:apam}.
\begin{table}
\begin{center}
\begin{tabular}{|c||c|c|c||c|c|c|}
\hline
$\alpha$ & $A^+$ & $c_1^+$ & $c_2^+$ & $A^-$ & $c_1^-$ & $c_2^-$  \\
\hline
 -0.007 & 0.3416(4) & 0.020(1)  &  -0.041(1)  
& 0.3317(4) & 0.048(1) & 0.086(1) \\
 -0.013 & 0.3636(6) & 0.022(1)  &  -0.049(2)
& 0.3445(6) & 0.085(1) & 0.161(2) \\
 -0.017 & 0.3790(8) & 0.015(1)  &  -0.041(3)
& 0.3533(8) & 0.109(2) & 0.211(4) \\
 -0.019 & 0.3870(9) & 0.010(2)  &  -0.033(4)  
& 0.3578(9) & 0.120(2) & 0.237(5) \\
 -0.025 & 0.4117(13) & -0.016(3)&  0.006(6)
& 0.3718(13) & 0.151(4) & 0.312(9) \\
\hline
\end{tabular}
\end{center}
\vspace{-0.2cm}
\caption{The parameters of the fits to Eq.\ (\ref{spfit}) for 
$\omega=0.79$ and some selected $\alpha$-values. The errors were
obtained by Monte Carlo variation of the parameters of $C_{ns}$
in Eq.\ (\ref{cnsfit}).}
\label{tab:cfitr}
\end{table}
A parametrization of the amplitudes as suggested by Eqs.\ (\ref{Aexp})
and (\ref{polcon}) 
\be
A^{\pm} \; = \; a_0 + a_1^{\pm}\alpha + a_2^{\pm}\alpha^2~,
\label{aparam}
\ee
works extremely well, as can be seen in Fig.\ \ref{fig:apam}, and confirms 
explicitly the cancellation of the pole terms as predicted in Eq.\ 
(\ref{polcon}). If $A^+$ and $A^-$ are independently fitted, that is 
with perhaps different $a_0$, we get $a_0^+=0.3176(12)$ and $a_0^-=
0.3175(12)$. The final result is found by using Eq.\ (\ref{aparam})
with fixed $a_0=0.3175$ (the error in $a_0=-c^p_{ns}$ is already included
in the errors of the $A^{\pm}$-values, which are now parametrized).
We obtain  
\ba
A^+ & \!\! = & \!\! a_0 -3.308(36) \alpha + 18.4(2.2) \alpha^2~,
\label{aplus}\\
A^- & \!\! = & \!\! a_0 -1.975(36) \alpha + \, ~7.8(2.2) \alpha^2~.
\label{aminus}
\ea
At this point it is appropriate to discuss the influence of an 
$\omega$-variation on $A^+$ and $A^-$. From Eq.\ (\ref{omvar}) we know
that a shift in $\omega$ of size $\Delta\omega=0.02$ shifts the pole 
term parameter $c^p_{ns}$ by about $0.3\%$ and therefore we expect
a shift of $a_0$ by the same amount. In fact that is exactly what 
happens and it is the only effect, because the new parameters $a_1^{\pm}$
and $a_2^{\pm}$ coincide inside error bars with the values found
for $\omega=0.79$. All in all that results in a common shift of the    
$A^+$ and $A^-$-curves in Fig.\ \ref{fig:apam} by again $0.3\%$.
As a consequence the universal amplitude ratio $A^+/A^-$ becomes 
essentially independent of $\omega$.
\begin{figure}[t!]
\begin{center}
   \epsfig{bbllx=63,bblly=265,bburx=516,bbury=588,
       file=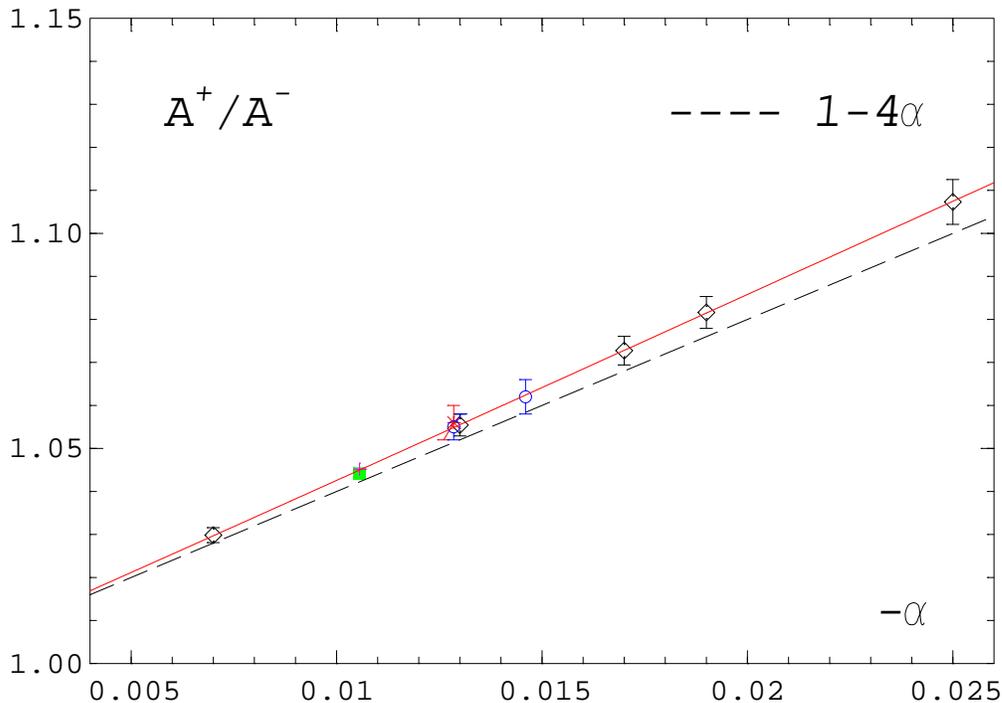, width=120mm}
\end{center}
\caption{The universal ratio $A^+/A^-$ versus $-\alpha$. The solid line
is obtained from Eqs.\ (\ref{aplus}) and (\ref{aminus}), the diamonds by
direct calculation from Table \ref{tab:cfitr}. The other symbols denote 
results from the shuttle experiment(square) \cite{Lipa:1996,Lipa:2000},
from Campostrini et al.\ (circles) \cite{Campostrini:2001zw,Campostrini:2001iw},
from Larin et al.\ (star) \cite{Larin:1998} and Kleinert et al.\
(plus) \cite{Kleinert:2000}.}
\label{fig:u0}
\end{figure}

The universal ratio $A^+/A^-$ is sometimes given in terms of a
function $\mathcal{P}(\alpha)$ \cite{Barmatz:1975} 
\be
A^+/A^- \; = \; 1 - \mathcal{P}\alpha~.
\label{Pfun}
\ee
Expanding the ratio in powers of $\alpha$ we arrive at the following 
relation for $\mathcal{P(\alpha)}$
\be
\mathcal{P} \; = \; {1 \over \alpha} \left ( 1- {A^+ \over A^-}\right )
\; = \; {a_1^- -a_1^+ \over a_0} + \left [ {a_2^- -a_2^+ \over a_0} 
-{a_1^- \over a_0}\cdot {a_1^- -a_1^+ \over a_0} \right ] \alpha +\dots~,
\label{Pexp}
\ee
that is, $\mathcal{P}$ goes to a finite limit when $\alpha \to 0$
\cite{Barmatz:1975,Griffiths:1967}. In fact, there is a phenomenological
relation \cite{Pelissetto:2000ek,Hohenberg:1976}
\be
A^+/A^- \; =\; 1 - 4\alpha~,
\label{pheno}
\ee
predicting $\mathcal{P}=4$. Evaluating Eqs.\ (\ref{aplus}) and 
(\ref{aminus}) leads to
\be
A^+/A^- \; =\; 1 - 4.20(5)\alpha + \dots~,
\label{u0}
\ee
rather close to the relation (\ref{pheno}).
In Fig.\ \ref{fig:u0} we show the ratio and compare it to former results
from the shuttle experiment \cite{Lipa:1996,Lipa:2000} as well as some
analytical determinations \cite{Campostrini:2001zw,Campostrini:2001iw}
and \cite{Larin:1998,Kleinert:2000}. We note that our ratio result is in
complete accordance with all of the other ratio results. Obviously, they
differ 
among each other simply and solely by assuming different $\alpha$-values.
This conclusion was already reached by Campostrini et al.\ 
\cite{Campostrini:2001zw}, we can however directly confirm it with 
Eqs.\ (\ref{aplus}) and (\ref{aminus}).

\section{$A^+/A^-$ from the Equation of State}
\label{section:eos}

\n The magnetic equation of state describes the critical behaviour of the
magnetization in the vicinity of $T_c$. As noted by Widom \cite{Widom:1965}
and Griffiths \cite{Griffiths:1967} already long ago the equation of state
may be integrated to yield the scaling function for the free energy. From
subsequent derivatives with respect to the temperature one obtaines then 
the specific heat and in particular an equation for the universal ratio 
$A^+/A^-$. Before we come to this relation we have to briefly discuss the
equation of state. The Widom-Griffiths form of the equation of state is 
given by
\be
y=f(x)~,
\label{wgeos}
\ee
where 
\be
y \equiv h/M^{\delta}~, \quad x \equiv \bar t/M^{1/\beta}~.
\label{yxdef}
\ee
The variables $\bar t$ and $h$ are the normalized reduced temperature and
magnetic field
\be
\bar t = (T-T_c)/T_0~, \quad h = H/H_0~,
\label{thnorm}
\ee
associated with the usual normalization conditions 
\be
f(0) = 1~, \quad {\rm and } \quad f(-1)=0~.
\label{fnorm}
\ee
The reduced temperature $\bar t$ differs from $t$ by a constant factor
($\bar t = [T_c/T_0]t$), because of the second condition in (\ref{fnorm}). 
The normalization constants can be expressed in terms of the critical
amplitudes from Eq.\ (\ref{camp})
\be
T_0 = B^{-1/\beta}T_c = 1.18(2)~, \quad H_0= D_c =1.11(1)~.
\label{nconst}
\ee
The numbers in the last equation have been obtained in Ref.\ 
\cite{Engels:2000xw} by assuming a special set \cite{Hasenbusch:1999}
of critical exponents
\be
\beta = 0.3490(6)~, \quad \nu = 0.6723(11)~,
\label{Hasex}
\ee
which implies $\alpha\approx -0.017$. The same is true for the equation
of state, which was determined numerically in \cite{Engels:2000xw} from
simulations with a non-zero magnetic field. Using this equation of state
will therefore give $A^+/A^-$ for only that particular value of $\alpha$.
Varying $\alpha$ in the range $[-0.0136,-0.0202]$, as suggested
by the error of $\nu$, would result in a large variation of $A^+/A^-$
to begin with (see Fig.\ \ref{fig:u0}). Insofar we consider the
following calculation mainly as a test of the method.

The results for the equation of state were parametrized in 
\cite{Engels:2000xw} by a combination of a small-$x$ (low temperature)
and a large-$x$ (high temperature) ansatz. The small-$x$
form $x_s(y)$ was inspired by perturbation theory 
\cite{Wallace:1975} and incorporates the divergence of the susceptibility
on the coexistence line ($x=-1;~ y=0$) due to the massless Goldstone modes 
\be 
x_s(y)+1 \;=\; ({\widetilde c_1} \,+\, {\widetilde d_3})\,y \,+\,
             {\widetilde c_2}\,y^{1/2} \,+\, 
             {\widetilde d_2}\,y^{3/2} \;.
\label{smallx}
\ee
The large-$x$ form $x_l(y)$ was derived from  
Griffiths's analyticity condition \cite{Griffiths:1967}
\be
x_l(y)\;=\; a\, y^{1/\gamma} \,+\, b\,y^{(1-2\beta)/\gamma}~.
\label{highx}
\ee  
The parameter values are
\ba
{\widetilde c_1} \,+\, {\widetilde d_3} &\!\! = \; ~0.352(30)~, \quad
~{\widetilde c_2} &\!\! = \; ~~0.592(10)~, \label{params} \\
a &\!\! = \; 1.2595(30)~, \quad  b &\!\! = \; -1.163(20)~.
\label{paraml}
\ea
Because of the normalization $y(0)=1$ we have ${\widetilde d_2}=
1-({\widetilde c_1}+{\widetilde d_3}+{\widetilde c_2})$. The complete
equation of state is obtained by interpolation of the low and high
temperature parts 
\be
x(y) \;=\; x_s(y)\,\frac{y_0^p}{y_0^p + y^p} \,+\,
           x_l(y)\,\frac{y^p}{y_0^p + y^p}~,
\label{totalfit}
\ee
with $p=6$ and $y_0 =3.5$.

For negative $\alpha$ the universal ratio $A^+/A^-$ can be calculated 
from $f(x)$ using the following formula \cite{Aharony:1974} 
\be
\frac{A^+}{A^-} \;=\; \frac{ - \int_0^{\infty} dx \, x^{\alpha -2} \,
                            [f'(0) - f'(x) + f''(0) x]}{
                     f'(0)/(1-\alpha) \,+\, f''(0)/\alpha \,+\,
              \int_{-1}^0 dx \, (-x)^{\alpha -2} \,
                            [f'(0) - f'(x) + f''(0) x]}
\;.
\label{ratio_int}
\ee
The main contribution to both the nominator and the denominator is 
$f''(0)/\alpha$. A more appropriate representation of $A^+/A^-$ is
therefore
\be
\frac{A^+}{A^-} \;=\; \frac{ 1 + [ \alpha / f''(0)]\, F_N}{
 1 + [\alpha / f''(0)]\, F_D}~,
\label{nearone}
\ee
where
\ba
\!\!\!\!F_N \!\!&\!\! =\!\! & \!\! -{ f'(0)\over 1-\alpha }
 - \int_0^1 dx \, x^{\alpha -2} \, [f'(0) - f'(x) + f''(0) x]
 + \int_1^{\infty} dx \, x^{\alpha -2} \, f'(x)\, ,
\label{FN} \\
\!\!\!\!F_D \!\!&\!\! =\!\!& \!\! { f'(0)\over 1-\alpha } \,+\,
\int^0_{-1} dx \, (-x)^{\alpha -2} \, [f'(0) - f'(x) + f''(0) x]\, .
\label{FD}
\ea
Let us denote the integrals in Eq.\ (\ref{FN}) by $I_1$ and $I_2$, 
the one in Eq.\ (\ref{FD}) by $I_3$. To a good approximation we can 
calculate the integrals $I_1$ and $I_3$ as well as the derivatives
from the low temperature equation (\ref{smallx}). In order to obtain 
$I_2$ we first rewrite the integral as 
\be
I_2\; = \; -f(1) + (2-\alpha) \int^{\infty}_{f(1)} dy y {dx \over dy}
x^{\alpha-3}~,
\label{I2}
\ee 
and evaluate the remaining integral from the interpolation formula
(\ref{totalfit}), using for $f(1)$ the low temperature value 2.4448.
For the derivatives we find
\ba
f'(0)\!\! &=&\!\! 2 \left( 3 - {\widetilde c_1} - {\widetilde d_3}
-2{\widetilde c_2} \right)^{-1} \;=\; 1.366 \pm 0.034~, \label{fp}\\
f''(0)\!\! &=&\!\! [f'(0)]^{\,3} \left( (3/4) ({\widetilde c_1} +
{\widetilde d_3} -1) +{\widetilde c_2} \right)  \;=\; 0.270 \pm 0.064~,
\label{fpp}
\ea
and for the integrals
\be
I_1 \;=\; 0.203\, \pm 0.02 , \quad I_2 \;=\; 1.749\,\pm 0.03 , \quad
I_3 \;=\; 0.512\, \pm 0.02 .  
\label{integrals} 
\ee
The errors in the integrals were obtained by Monte Carlo variation of the 
initial parameters in Eqs.\ (\ref{params}) and (\ref{paraml}). When
this procedure is also applied to the complete expression (\ref{nearone})
one obtains 
\be
A^+/A^- \; = \; 1.12 \pm 0.05~.
\label{ratint}
\ee
The first conclusion to be drawn from this result is that this method is
not well suited for the calculation of the ratio, at least with the 
parametrization of the equation of state of Ref.\ \cite{Engels:2000xw}.
Though the result (\ref{ratint}) is compatible with our directly 
determined ratio $A^+/A^-(\alpha=-0.017)\, =\, 1.073(3)$, the error is
rather large. The main source of the error is evidently the inaccurate
value of $f''(0)$. That this quantity plays an important role is of course
not unexpected, because $A^+$ and $A^-$ are the amplitudes of the specific
heat, which is again the second derivative of the free energy density.
Our parametrization was not devised for that purpose, but for a correct 
description of the Goldstone effect near to $x=-1$ and the limiting 
behaviour for $x \to \infty$. That is why it led to a precise
determination of $R_{\chi}$ and the constant $c_f$
\be
R_{\chi} = \lim \limits_{x \to \infty} x^{\gamma}/f(x)\, =\, 1.356(4)\, ,
\quad c_f \equiv \lim \limits_{x \to -1} (1+x)^{-2}f(x)\, =\, 2.85(7)\,.
\label{eqconst}
\ee
\indent Campostrini et al.\ have used a different representation of the equation
of state \cite{Campostrini:2000si,Campostrini:2001iw}, based on Josephson's
parametrization \cite{Josephson:1969} of $M,\bar t$ and $H$ in terms of the
variables $R$ and $\theta$ and parametric functions.
 In order to fix these functions approximately the authors
utilized the results of an analysis of the high-temperature expansion of an
improved lattice Hamiltonian. The values obtained for $A^+/A^-$ compare well
with our direct determination and were already shown in Fig.\ \ref{fig:u0}.
The corresponding equation of state differs however somewhat in the low and
medium temperature regions from the data points from our non-zero field 
simulations \cite{Engels:2000xw}. The question arises then whether the same
data may be described as well in the schemes introduced by Campostrini et
al.\ . Such alternative fits of the data have been carried out by two of us
\cite{Cucchieri:2001}. The $\chi^2$ per degree of freedom of these fits is
generally high, in particular for scheme $A$ of Ref.\ 
\cite{Campostrini:2000si}. The fits according to scheme $B$ are
considerably better and lead to a ratio $A^+/A^- = 1.070(13)$, again 
compatible with our direct determination. The simultaneously calculated
ratio $R_C$ is however much larger (0.165-0.185) than expected from
analytical calculations (0.123-0.130) \cite{Strosser:1999pt,Kleinert:2000}.  
We therefore do not pursue this method of calculation here in more detail.
%
\section{Simulations with $H>0$}
\label{section:simh}
\n We have performed additional simulations with a positive magnetic field
$H$ on the critical line to find the remaining critical amplitudes for the
specific heat and the longitudinal and transverse correlation lengths.
\begin{table}
\begin{center}
\begin{tabular}{|r|c|c|c|c||c|}
\hline
$L$ & $H$-range & $N_{cu}$ & $N_{meas} [1000]$ & $N_H$ &  $N_{tot}$ \\
\hline
   36 & 0.0007-0.05 & 50-100 &   30-40  & 25  & 36  \\
   48 & 0.0001-0.03 & 50-100 &   30-40  & 30  & 39  \\
   72 & 0.0001-0.005 & 60-300 &  20     & 15  & 23  \\
   96 & 0.0001-0.0015 & 60-80 & 12-20 & 8 & 16\\
\hline
\end{tabular}
\end{center}
\caption{Survey of the new Monte Carlo simulations at $T_c$ on different
lattices. $N_{cu}$ is the number of cluster updates between the 
measurements, $N_{meas}$ the number of measurements per $H$-value in 
units of 1000 and $N_H$ the number of $H$-values at which new
runs were performed. $N_{tot}$ is the total number of $H$-values where we
have data.}
\vspace*{-0.2cm}
\label{tab:nonezero}
\end{table}
The linear extensions of the lattices we used were $L = 36,48,72$ and 96.
These measurements were combined with those from Ref.\ \cite{Engels:2000xw}
to cover the $H$-range appropriately. Some of the new data have already 
been used in Ref.\ \cite{Engels:2001bq}. In Table \ref{tab:nonezero}
we give more details of these simulations. 
%
\subsection{The Specific Heat on the Critical Line}
\label{section:Ac}
\begin{figure}[b]
\begin{center}
   \epsfig{bbllx=63,bblly=265,bburx=516,bbury=588,
       file=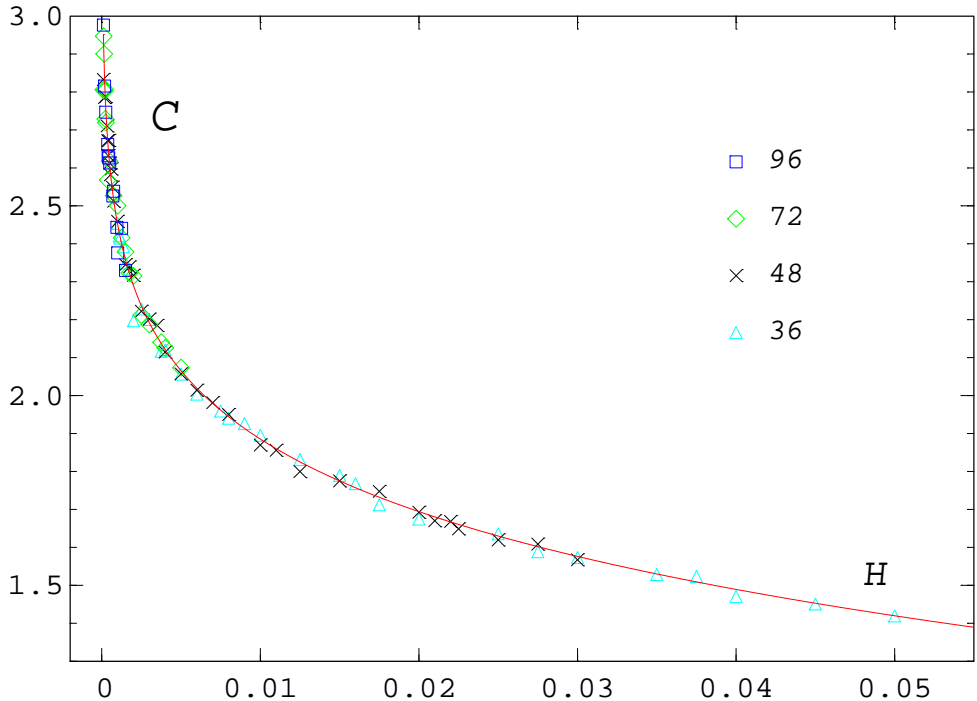, width=120mm}
\end{center}
\caption{The specific heat at $T_c$ for $L= 36,48,72$ and 96 as a 
function of $H$. The line is the fit (\ref{fitch}) for $\alpha_c = -0.0078\,
(\nu=0.671)$ and $\omega=0.79$.}
\label{fig:ch}
\end{figure}

\begin{figure}[ht]
\vspace*{0.2cm}
\begin{center}
   \epsfig{bbllx=63,bblly=280,bburx=516,bbury=563,
       file=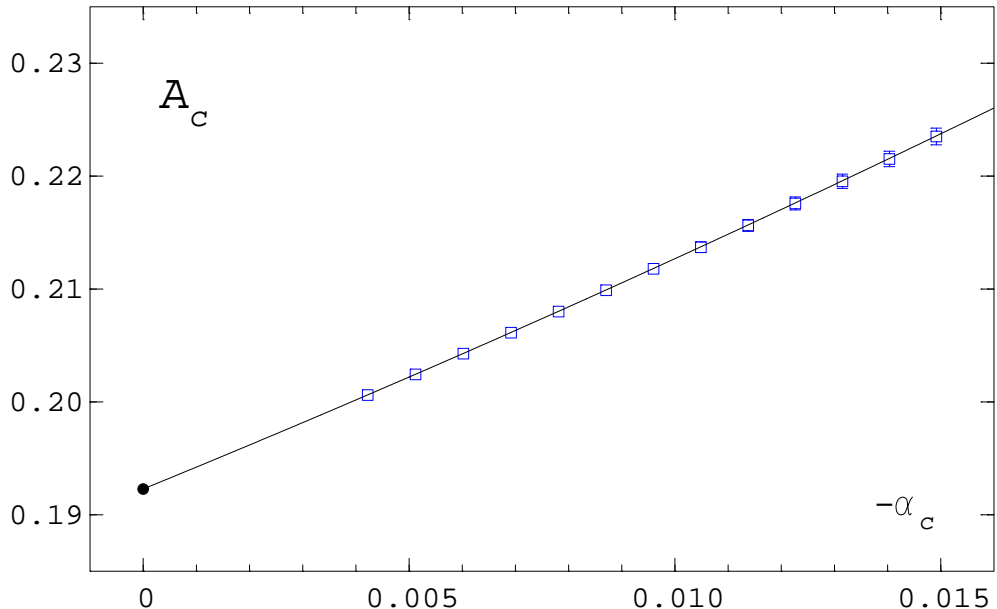, width=120mm}
\end{center}
\caption{The amplitude $A_c$ versus $-\alpha_c$ (squares) for $\omega=0.79$.
The filled circle shows the value expected from $C_{ns}$, the line is the 
parametrization (\ref{acalc}).}
\label{fig:ac}
\end{figure}

%
In Fig.\ \ref{fig:ch} we show our specific heat data as a function of the
magnetic field $H$. Since there are no noticeable systematic finite size 
effects we can use these data to fit them to the ansatz
\be
C \;=\; C_{ns} + {A_c \over \alpha_c} H^{-\alpha_c} (1+
c_h H^{\omega\nu_c}) ~.
\label{fitch}
\ee
Here, $C_{ns}$ is the same non-singular term, which we have already 
determined in Section \ref{section:sphattc} as a function of $\alpha$
(or $\nu$) with the result (\ref{cnsfit}). Because of the dependence of
$C$ on $\alpha_c$ and $\nu_c$ the amplitudes $A_c$ and $c_h$ depend
on two critical exponents. The second exponent will however not introduce
a sizeable variation in the amplitudes. We therefore treat the exponent
$\beta$ as fixed to the value $\beta =0.349\,$, in accord with our previous
calculations. With the relations
\be
\beta \delta \; = \; 2-\beta -\alpha~,\quad
\alpha_c \; = \; {\alpha \over  2-\beta -\alpha}~,\quad
\alpha  \; = \; {\alpha_c (2-\beta) \over 1+ \alpha_c}~,
\label{aac}
\ee
the linear dependence of $C_{ns}$ on $1/\alpha$ can be rewritten as one on
$1/\alpha_c$
\ba
C_{ns} &=& c_{ns}^0 +{ c_{ns}^p \over 2- \beta} \left(1 + 
{1 \over \alpha_c} \right) \vspace*{0.2cm} \\
       &=& 3.16(4) - {0.1923(3) \over \alpha_c}~.
\label{cnsac}
\ea
\begin{table}[t]
\begin{center}
\begin{tabular}{|c|c||c|c||c|}
\hline
$\alpha_c$ & $\alpha$ & $A_c$ & $c_h$ & $\chi^2/N_f$ \\
\hline
-0.00422 & -0.007 & 0.2006(2) & 0.0203(1)  &  1.09 \\
-0.00781 & -0.013 & 0.2080(3) & 0.0344(2)  &  1.09 \\
-0.01019 & -0.017 & 0.2131(5) & 0.0423(4)  &  1.10 \\
-0.01138 & -0.019 & 0.2156(5) & 0.0458(4)  &  1.10 \\
-0.01492 & -0.025 & 0.2235(7) & 0.0546(8)  &  1.11 \\
\hline
\end{tabular}
\end{center}
\vspace{-0.2cm}
\caption{The parameters of the fits to Eq.\ (\ref{fitch}) for some 
selected $\alpha_c$-values at fixed $\beta=0.349$ and $\omega=0.79$.
The errors were obtained by Monte Carlo variation of the parameters of 
$C_{ns}$ in Eq.\ (\ref{cnsac}).}
\label{tab:ac}
\end{table}
We took this form of $C_{ns}$ as an input to the fits of $C$ with Eq.\
(\ref{fitch}). The $H$-range for the fits was $0.0001 \le H \le 0.05$. We
have convinced ourselves that smaller $H$-ranges (up to 0.02 or 0.03)
lead inside the error bars to the same results for the amplitudes. In
Table \ref{tab:ac} we present details of the fits for several 
$\alpha_c$-values, in Fig.\ \ref{fig:ac} we show the amplitude $A_c$ as
a function of $\alpha_c$. As in the case of the amplitudes $A^{\pm}$ the
pole of $C_{ns}$ in Eq.\ (\ref{cnsac}) is compensated by the corresponding
pole term in $A_c/\alpha_c$. We have therefore parametrized the 
$\alpha_c$-dependence
of $A_c$ in analogy to Eq.\ (\ref{aparam}) with the fixed value 
$A_c(\alpha_c=0)\,=\, 0.1923\,$ and find
\be
A_c \; =\; 0.1923 -1.919(42)\alpha_c +11.6(4.1) \alpha_c^2~.
\label{acalc}
\ee
From Fig.\ \ref{fig:ac} we see that this parametrization describes the data
very well. Like in the study of the $\omega$-dependence of $A^{\pm}$ in 
Section \ref{section:heatratio} we found changes of similar size for
the amplitude $A_c$ due to a variation of $\omega$. They lead to an
additional error of $A_c$ of size 0.0006 at $\alpha_c=-0.00422$, 
which decreases to 0.0004 at $\alpha_c=-0.01492$.  

%
\subsection{The Correlation Lengths on the Critical Line}
\begin{figure}[ht]
\begin{center}
   \epsfig{bbllx=63,bblly=280,bburx=516,bbury=563,
       file=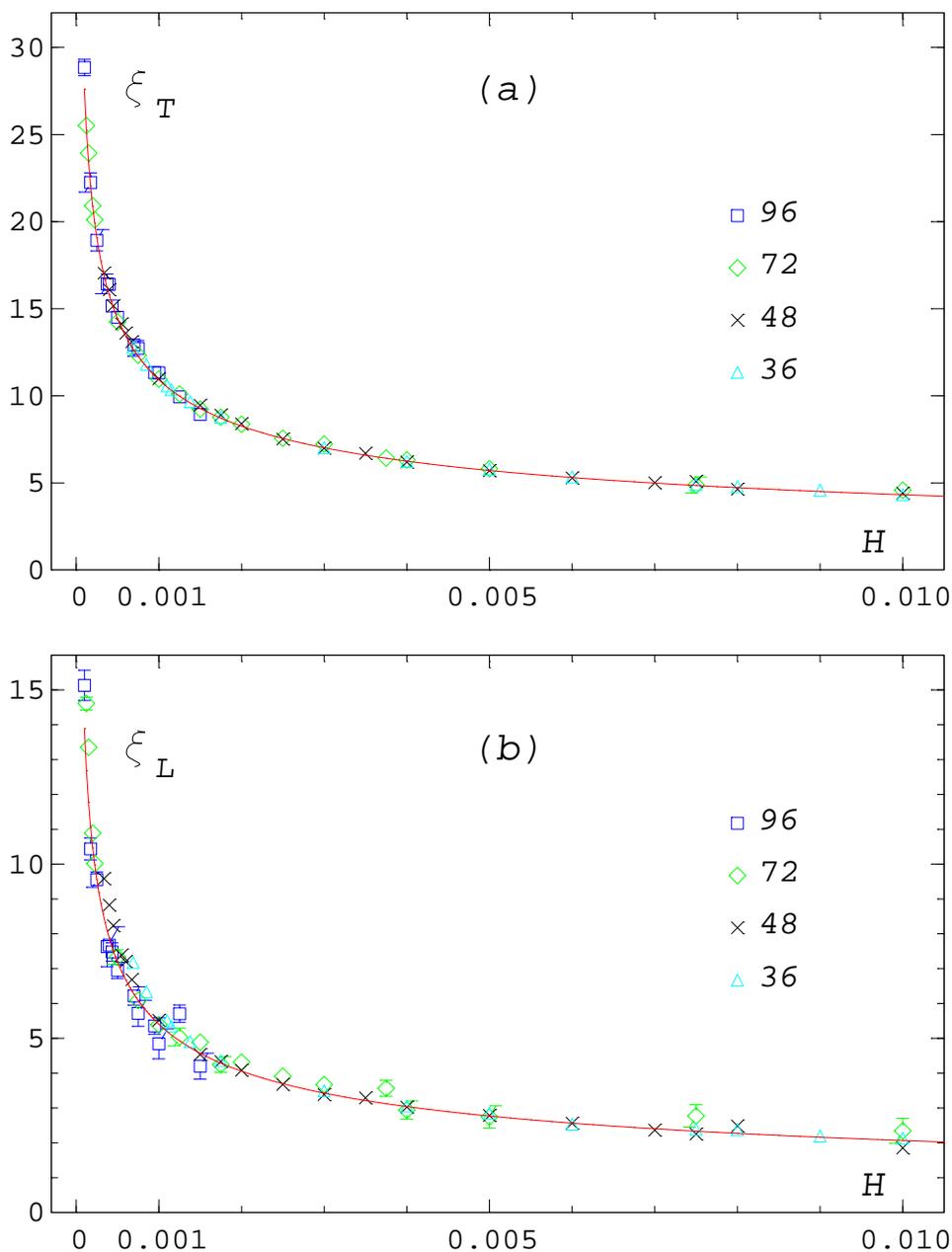, width=120mm}
\end{center}
\caption{The correlation lengths $\xi_T$ (a) and $\xi_L$ (b) at $T_c$ 
for $L= 36,48,72$ and 96 as a function of $H$. The lines are the fits 
(\ref{fitcor}) for $\nu_c =0.40325$ and $\omega=0.79\,$.}
\label{fig:cole}
\end{figure}

%
\n The simulation results for the transverse and longitudinal correlation
lengths are shown in Fig.\ \ref{fig:cole}$\;$a) and b). For the transverse 
correlation length $\xi_T$ one can hardly detect finite size effects,
whereas the longitudinal correlation length $\xi_L$ shows more fluctuations
and a systematic deviation to higher $\xi_L$-values, when one decreases 
the magnetic field $H$. The smaller the lattice, the earlier this behaviour
sets in. In order to determine the amplitudes we have fitted our results
to the following form
\be
\xi_{T,L} \; = \;\xi_{T,L}^c H^{-\nu_c} \left( 1 + c_{T,L} H^{\omega\nu_c}
\right)~.  
\label{fitcor}
\ee 
\begin{table}[b]
\begin{center}
\begin{tabular}{|c|c||c|c||c|c|}
\hline
$\nu_c$ & $\alpha$ & $\xi_T^c$ & $c_T$ & $\xi_L^c$ & $c_L$ \\
\hline
0.40350 & -0.007 & 0.6709(14) & 0.024(13)  & 0.3427(15)  
& -0.258(33) \\
0.40325 & -0.013 & 0.6724(14) & 0.019(14)  & 0.3435(15)
& -0.263(33) \\
0.40307 &  -0.017 & 0.6735(14) & 0.015(14)  & 0.3441(15)
& -0.266(33) \\
0.40299 & -0.019 & 0.6740(14) & 0.013(14)  & 0.3443(15)  
& -0.268(32) \\
0.40274 & -0.025 & 0.6755(14) & 0.008(14)  & 0.3451(15)
& -0.273(32) \\
\hline
\end{tabular}
\end{center}
\vspace{-0.2cm}
\caption{The parameters of the fits to Eq.\ (\ref{fitcor}) for some 
selected $\nu_c$-values and $\omega=0.79$. The $\chi^2/N_f$-values
varied for $\xi_T$ between 0.89 and 0.86, for $\xi_L$ it was $0.67$.}
\label{tab:xicri}
\end{table}
\newpage
\begin{figure}[t!]
\begin{center}
   \epsfig{bbllx=63,bblly=280,bburx=516,bbury=563,
       file=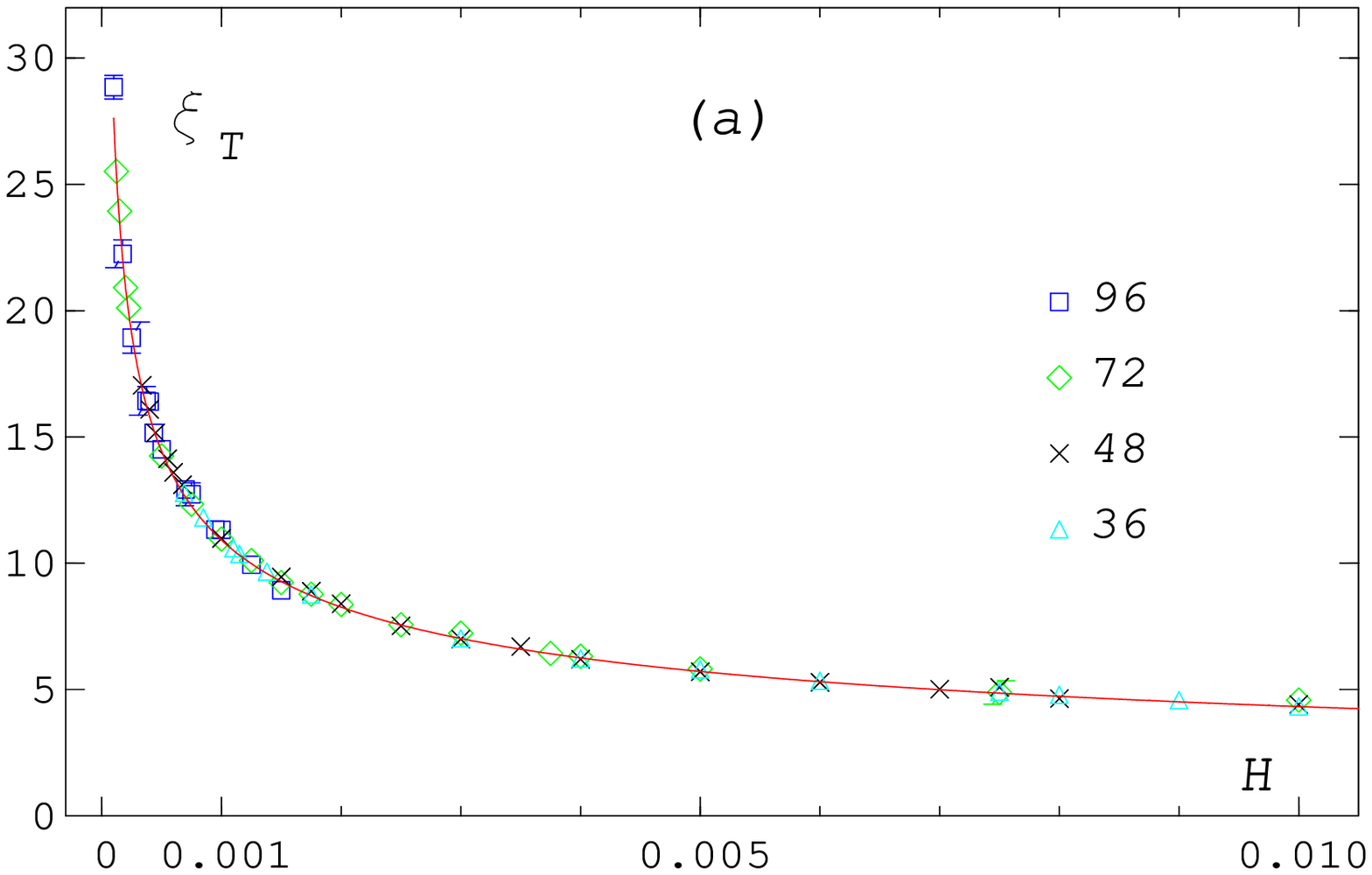, width=120mm}
\end{center}
\end{figure}
\n In the transverse case we used the reweighted data for $L=72$ in the 
$H$-interval [0.0005,0.0025], for $L=48$ in [0.002,0.02] and for $L=36$
in [0.015,0.03]. From Table \ref{tab:xicri} we see that the correction
term is essentially zero. Correspondingly, there is no $\omega$-dependence
and a fit with $c_T\equiv 0$ works
just as well (even with the same $\chi^2/N_f$), and leads to a slight 
increase in the amplitude value, which is of the order of the error given
in Table \ref{tab:xicri}. The dependence of the amplitude $\xi_T^c$ on 
$\nu_c$ or $\alpha$ is linear but the slope is very small. In order to
determine the longitudinal amplitude $\xi_L^c$ we have fitted the 
reweighted data for $L=72$ in the $H$-interval [0.0005,0.00175] together 
with those for $L=48$ in [0.00175,0.01]. Here, the correction term is not
zero, but the variation due to $\omega$ is still negligible. The $\nu_c$- or 
$\alpha$-dependence is the same as for $\xi_T^c$,
the ratio of the two correlations lengths is a fixed number
\be
\xi_T^c/\xi_L^c \; =\; 1.957(10)~,
\label{ratxi}
\ee
independent of the critical exponents. It is well-known (see Refs.\
\cite{Privman:1991} and \cite{Patashinskii:1973,Fisher:1985}) that
at zero field on the coexistence line $t<0$ the longitudinal correlation  
function $G_L$ is for large distances $|\vec r|$ connected to the transverse
one by  
\be
G_L (\vec r ,t )\; \approx \; {1 \over 2}(N-1)[G_T (\vec r ,t )/M]^2~,
\label{corfunc}
\ee
where in our case $N=2$.
The relation is expected to hold also for small non-zero fields $H$ near
the phase boundary in the regime of exponential decay implying a factor 2   
between the correlation lengths. It is remarkable, that we find 
approximately such a value for the ratio at $t=0$. A similar observation 
has been made for the $3d~O(4)$ model \cite{Engels:2002}.
%
\subsection{The Stiffness Constant on the Coexistence Line}
%
The stiffness constant $\rho_s(T)$ is related to the helicity modulus
$\Upsilon$ \cite{Fisher:1973} by
\be
\rho_s \; =\; \Upsilon / T~,
\label{stiffmod}
\ee
which can be measured in Monte Carlo simulations. This was done e.\ g.\
in Refs.\ \cite{Li:1989} and \cite{Janke:1990}. Here we follow a different
strategy, which we applied already in Ref.\ \cite{Engels:2000xw} to find
the magnetization on the coexistence line. The $L$ or volume dependence
of $M$ at fixed $J$ and fixed small $H$ is described by the 
$\epsilon$-expansion of chiral perturbation theory (CPT) in terms of 
two low energy constants. One is the Goldstone-boson-decay constant $F$,
the other the magnetization $\Sigma$ of the continuum theory for $H=0$
and $V\rightarrow \infty$. The square of the constant $F$ is proportional
to the helicity modulus. In our notation, which is different from the one
in CPT (see the remark in the last paragraph of Ref.\ \cite{Tominaga:1994bc})
we have
\be
\Upsilon \; =\; F^2/J~, \quad {\rm implying}\quad \rho_s \; =\; F^2~.
\label{sm2}
\ee
The formulae, which are needed for the fits to determine the constants,
are summarized in Ref.\ \cite{Engels:2000xw} and were taken from Ref.\ 
\cite{Tominaga:1994bc}. In Table \ref{tab:ffit} we list the results for
the Goldstone-boson-decay constant $F$ at various $J$-values. We 
performed simulations at $H=0.0001$ on lattices with linear extensions
$L=8,10,12,16,20,24,30,36,40,48$ and 56. By construction the 
$\epsilon$-expansion is only applicable in a range where $m_{\pi}L\lsim 1$.
This condition translates into the equation 
\be
H{\Sigma \over \sqrt{J}}\; \lsim \;\left({F \over L}\right)^2~,
\label{cond}
\ee
\begin{table}[t]
\begin{center}
\begin{tabular}{|c||c|c||c|c|}
\hline
$J=1/T$ & $F$ & $\Delta F$ & $L_{min}$ & $L_{max}$   \\
\hline
 0.462   & 0.1993 & 0.0096 & 8,10,12  & 36,40 \\
 0.465   & 0.2275 & 0.0060 & 8,10,12  & 40,48 \\
 0.470   & 0.2596 & 0.0050 & 8,10,12  & 40,48 \\
 0.480   & 0.3091 & 0.0018 & 8,10,12  &  48   \\
 0.500   & 0.3795 & 0.0114 & 8,10,12  & 48,56 \\
 0.525   & 0.4379 & 0.0040 & 8,10,12  & 48,56 \\
 0.550   & 0.4755 & 0.0028 & 8,10,12  &  56   \\
\hline
\end{tabular}
\end{center}
\vspace{-0.2cm}
\caption{The Goldstone-boson-decay constant $F$ at various $J$-values
from fits on data from lattices with $L$ in the range
$[L_{min},L_{max}]$.}
\label{tab:ffit}
\end{table}
\begin{figure}[hb]
\begin{center}
   \epsfig{bbllx=63,bblly=265,bburx=516,bbury=588,
       file=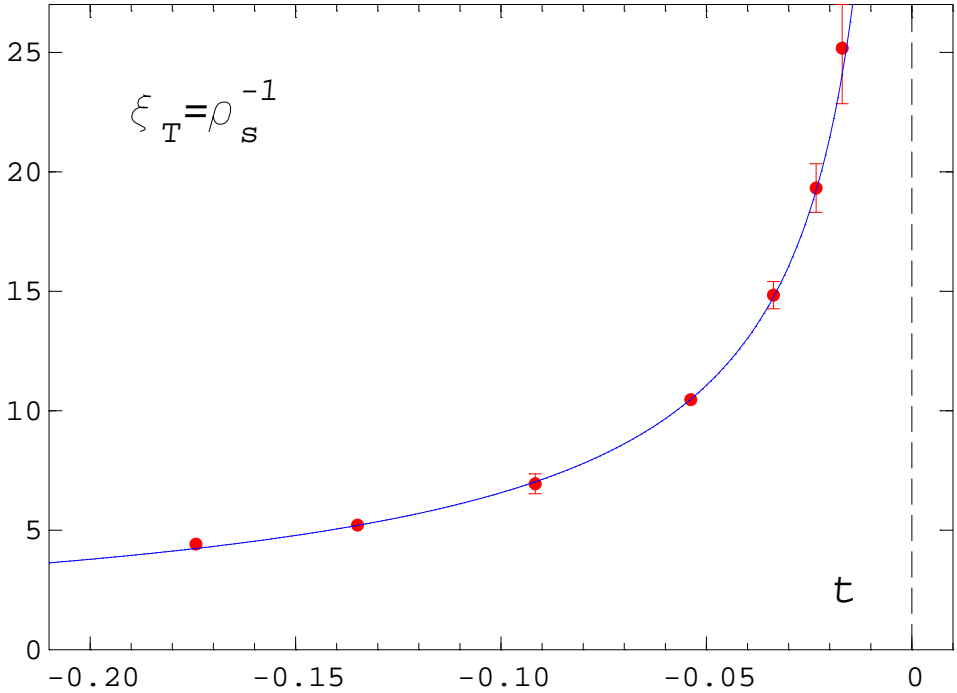, width=120mm}
\end{center}
\caption{The inverse of the stiffness constant $\rho_s^{-1}=\xi_T$
on the coexistence line from chiral perturbation theory. The solid line 
is the fit (\ref{fitxim}) with $\nu=0.671$ and $\omega=0.79$.}
\label{fig:stiff}
\end{figure}
\n and excludes the use of too large $L$-values. For each $J$ we fitted
different sets of data from lattices between $[L_{min},L_{max}]$ and 
averaged the obtained $F$-values. The errors on $F$ include the variations 
of these results. If we compare our $F$-values to the corresponding ones
of Ref.\ \cite{Tominaga:1994bc} we find generally somewhat lower numbers.
This may be due to the fact that in Ref.\ \cite{Tominaga:1994bc} data from
single lattices instead of sets of data from different lattices were fitted.
The transverse correlation length $\xi_T$ on the coexistence line is now
derived from the inverse of the stiffness constant or $F^{-2}$. It is 
plotted in Fig.\ \ref{fig:stiff}. Here, we have not as many and as accurate 
data as in Fig.\ \ref{fig:cole} a). In order to determine the amplitude
$\xi_T^-$ we fit our data points up to $J=0.525$ to the ansatz
\be
\xi_T\; = \; \xi_T^- (-t)^{-\nu}\left( 1+ c_T^-(-t)^{\omega\nu}\right)~.
\label{fitxim}
\ee
Table \ref{tab:xitm} contains the fit parameters for different $\nu$ or
$\alpha$-values. We observe, as for $\xi_T^c$, a linear dependence of 
the amplitude $\xi_T^-$ on $\alpha$ with a very small slope. A change of 
$\omega$ by 0.02 leads only to a shift in $\xi_T^-$ of a tenth of the error
in Table \ref{tab:xitm}.
\begin{table}[h]
\begin{center}
\begin{tabular}{|c|c||c|c||c|}
\hline
$\nu$ & $\alpha$ & $\xi_T^-$ & $c_T^-$ & $\chi^2/N_f$ \\
\hline
0.6690  & -0.007 & 1.680(52) & -0.55(10)  &  0.08 \\
0.6710  & -0.013 & 1.665(52) & -0.54(11)  &  0.08 \\
0.6723  & -0.017 & 1.655(51) & -0.53(11)  &  0.08 \\
0.6730  & -0.019 & 1.650(51) & -0.53(11)  &  0.08 \\
0.6750  & -0.025 & 1.636(51) & -0.52(11)  &  0.07 \\
\hline
\end{tabular}
\end{center}
\vspace{-0.2cm}
\caption{The parameters of the fits to Eq.\ (\ref{fitxim}) for several 
$\nu$-values and $\omega=0.79$.}
\label{tab:xitm}
\end{table}
%
\section{The Universal Amplitude Ratios}
\label{section:ratios}
%
After having determined all the amplitudes which appear in Eqs.\ (\ref{uratios})
to (\ref{hratios}) we can calculate the corresponding universal ratios. Since
the ratio $U_0=A^+/A^-$ has already been discussed in great detail we start with
the ratio $U_{\xi}$ of the correlation lengths for $H=0$. From 
Eq.\ (\ref{xiplus}) and Table \ref{tab:xitm} we find
\be
U_{\xi}\; = \; \xi^+ /\xi^-_T \; =\; 0.293(9)~,
\label{uxi}
\ee
independent of the used $\alpha$-value. The $\epsilon$-expansion of this 
ratio was derived by Hohenberg et al.\ \cite{Hohenberg:1976} to
$O(\epsilon)$ and extended by Bervillier \cite{Bervillier:1976ie} to
$O(\epsilon^2)$ resulting in $U_{\xi} = 0.27$ and 0.33, respectively.
Okabe and Ideura \cite{Okabe:1981} corrected the expansion of Bervillier
(not the numerical value) and computed the ratio in $1/N$-expansion to
$U_{\xi} = 0.140$. The $\epsilon$-expansion results are comparable in 
size to our value in (\ref{uxi}), the $1/N$-expansion result, however,
seems to be too small.
 
The ratios connecting the specific heat and correlation length amplitudes
are related by
\be
R_{\xi}^+ \; = \; R_{\xi}^T {U_0}^{1/d} U_{\xi}~,
\label{rcon}
\ee
and they depend on the used $\alpha$, mainly because of the specific heat
amplitudes. In Table \ref{tab:ratios} we have listed the ratios $R_{\xi}^+$
and $R_{\xi}^T$. From the $\alpha$-expansions (\ref{xiplus}) and 
(\ref{aplus}) we find 
\be
R_{\xi}^+ \; = \; 0.3382(14) -0.717(96) \alpha +0.87(1.13) \alpha^2~.
\label{rxia}
\ee
For $R_{\xi}^T$ one can derive a similar formula representing the values
of Table \ref{tab:ratios}
\be
R_{\xi}^T \; = \; 1.1580 -0.696 \alpha +0.97 \alpha^2 \pm 0.036~.
\label{rxita}
\ee
There exist several theoretical estimates of $R_{\xi}^+$ which compare 
well with our result: 0.355(3) [$\alpha=-0.0146$] 
\cite{Campostrini:2001iw} and 0.361(4)
\cite{Butera:1999qa}, both from high-temperature expansions; 0.36
\cite{Bervillier:1976ie} from the $\epsilon$-expansion, and 0.3597(10)
\cite{Bervillier:1980} and 0.3606(20) \cite{Bagnuls:1985} from $3d$ 
field theory. Apart from the first result, we could not relate a definite
$\alpha$-value to the respective estimate. The ratio $R_{\xi}^T$ was 
calculated from the $\epsilon$-expansion \cite{Hohenberg:1976,
Bervillier:1976ie} with the result 1.0(2) \cite{Privman:1991}, well in
accord with our value. 
\begin{table}[t]
\begin{center}
\begin{tabular}{|c|c||c|c||c||c|c|}
\hline
$\nu$ & $\alpha$ & $R_{\xi}^+$ & $R_{\xi}^T$ & $R_C$ & $R_A$ & $Q_2^T$\\
\hline
0.6690 & -0.007 & 0.3432(15) & 1.163(36) & 0.118(4) & 0.0515(17) & 0.834(21)\\
0.6710 & -0.013 & 0.3476(18) & 1.167(36) & 0.125(4) & 0.0534(18) & 0.849(21)\\
0.6723 & -0.017 & 0.3505(21) & 1.170(36) & 0.130(4) & 0.0547(18) & 0.860(21)\\
0.6730 & -0.019 & 0.3520(22) & 1.171(36) & 0.133(5) & 0.0554(19) & 0.865(21)\\
0.6750 & -0.025 & 0.3563(27) & 1.176(36) & 0.142(5) & 0.0574(19) & 0.881(22)\\
\hline
\end{tabular}
\end{center}
\vspace{-0.2cm}
\caption{The universal ratios from Eqs.\ (\ref{rxiratios}), (\ref{rce}) and
(\ref{hratios}) as a function of the used exponents $\nu$ and $\alpha$.}
\label{tab:ratios}
\end{table}
  
The remaining universal ratios $R_{\chi}, R_C, R_A$ and $Q_2^T$ are all
dependent on the amplitude $C^+$ of the susceptibility and/or the amplitudes
$B$ and $d_c(D_c)$ of the magnetization. We mentioned already that we had
determined $R_{\chi}, B$ and $d_c$ in Ref.\ \cite{Engels:2000xw}, although
for fixed $\nu=0.6723$. In the following we proceed as in Section 
\ref{section:Ac}, that is we keep $\beta$ fixed to 0.349 and assume in
addition that the $\nu$-dependencies of $R_{\chi}, B$ and $d_c$ are negligible.
In Table \ref{tab:ratios} we present the ratios $R_C$ and $Q_2^T$ as
calculated from
\be
R_C \;=\; A^+ R_{\chi} D_c^{-1} B^{-1-\delta}~,\quad
Q_2^T \;=\; (\xi^c_T / \xi^+)^{\gamma/\nu} R_{\chi} (d_c/B)^{\delta -1}
/(1+1/\delta)~,
\label{newrat}
\ee
and $R_A$ directly from the definition in Eq.\ (\ref{hratios}), using our
newly determined amplitudes $A^+, A_c, \xi^c_T$ and $\xi^+$. We could not find
any previous results for $R_A$ and $Q_2^T$ in the literature, however, the
ratio $R_C$ has been calculated theoretically in several ways. From Table
\ref{tab:ratios} we see that $R_C$ is increasing with decreasing $\alpha$,
which is due to the factor $A^+$. In comparing our values to the analytical
results we quote therefore the used $\alpha$-values. The ratio $R_C$
calculated from $3d$ field theory in Ref.\ \cite{Strosser:1999pt} is 
0.123(3) [$\alpha=-0.01285$], in Ref.\ \cite{Kleinert:2000} 
0.12428 [$\alpha\!=\!-0.01056$]; from the high-temperature expansion in Ref.\
\cite{Campostrini:2001iw} one finds 0.127(6) [$\alpha=-0.0146$]. The
results are in full agreement with our calculation, though that of Ref.\
\cite{Kleinert:2000} is somewhat higher than the other ones.
The old $\epsilon$-expansion result 0.103 of Aharony and Hohenberg 
\cite{Aharony:1976} seems to be too small.

\section{Conclusions}
\label{section:conclusion}
We have calculated the major universal amplitude ratios of the 
three-dimensional $O(2)$ model from Monte Carlo simulations. To reach this
goal a large amount of computer time had to be spent on the cluster of
alpha-workstations of the department of physics at the University of Bielefeld.
Most of the computer time went into the production of reliable specific heat
data for the direct determination of $A^+/A^-$. Initially we had hoped to 
improve the accuracy of the exponent $\alpha$ (or $\nu$) from these data. As
it turned out, however, the specific heat data could be fitted to a whole 
range of $\alpha$-values with the same $\chi^2/N_f$, extending even to 
$\alpha=0$. This raises the question, whether the experimental shuttle data
are really fixing the $\alpha$-value to exactly -0.01056, the same value as
in $3d$ field theory expansions \cite{Zinn-Justin:2001bf}. The positive
aspect of the indifference of the fits to the specific heat data to 
$\alpha$-variations was that we could study the numerical changes induced 
by these variations in the universal ratio $A^+/A^-$ and the background 
term $C_{ns}$. As a result we were able to confirm the conjectured pole
(in $1/\alpha$) behaviour of the amplitudes and the background term and
the mutual cancellation of the pole contributions. The same pole behaviour
was observed for the specific heat amplitude on the critical line. The
functional dependence of $A^+/A^-$ on the used $\alpha$-value is in complete 
accordance with all other ratio results and not far from the phenomenological
relation $A^+/A^- = 1-4\alpha$. We have also determined $A^+/A^-$ from 
the numerical equation of state, but we think the method relies too much
on the chosen parametrization. 

In order to find the amplitude of the transverse correlation length on the
coexistence line we used chiral perturbation theory. This enabled us to
calculate the less known ratios $R_{\xi}^T$ and $U_{\xi}$. The latter is
independent of the used $\alpha$, like the ratio $\xi_T^c/\xi_L^c$ on the critical
line, which is remarkably close to 2 - a prediction expected for $T<T_c$ 
from the correlation functions close to the phase boundary. Our results
for $R_{\xi}^+$ and $R_C$ are in full agreement with the best theoretical
estimates; $R_A$ and $Q_2^T$ are new and remain untested for the moment. 
\vskip 0.2truecm
\noindent{\Large{\bf Acknowledgements}}


\n We are grateful to Jean Zinn-Justin, Michele Caselle, Martin Hasenbusch
and Andrea Pelissetto for discussions and to Ettore Vicari for his comments
on the calculation of $A^+/A^-$ from the equation of state. Our work was
supported by the Deutsche Forschungs\-ge\-meinschaft under Grant No.\
FOR 339/1-2, the work of A.~C. and T.~M. in addition by FAPESP, Brazil
(Project No. 00/05047-5).

\clearpage

\end{document}